\documentclass[aps,pra,twocolumn]{revtex4-1}
\usepackage{amsmath,amsthm,amsfonts,amssymb}
\usepackage{xcolor,bm}
\usepackage{mathrsfs}
\usepackage{graphicx}
\usepackage{dsfont}
\usepackage{hyperref}
\usepackage{braket}
\usepackage{todonotes}
\usepackage{algorithm, algpseudocode}
\usepackage{algorithmicx}
\newtheorem{theorem}{Theorem}

\newtheorem{corollary}{Corollary}
\newtheorem{lemma}{Lemma}
\newtheorem{definition}{Definition}
\newtheorem*{remark*}{Remark}

\begin{document}
\title{Topological graph states and quantum error correction codes}

\author{Pengcheng Liao}
\affiliation{Institute for Quantum Science and Technology and Department of
Physics and Astronomy, University of Calgary, Calgary, Alberta T2N 1N4, Canada}
\author{Barry C. Sanders}
\affiliation{Institute for Quantum Science and Technology and Department of
Physics and Astronomy, University of Calgary, Calgary, Alberta T2N 1N4, Canada}
\author{David L. Feder}
\affiliation{Institute for Quantum Science and Technology and Department of
Physics and Astronomy, University of Calgary, Calgary, Alberta T2N 1N4, Canada}

\begin{abstract}
Deciding if a given family of quantum states is topologically ordered is an 
important but nontrivial problem in condensed matter physics and quantum 
information theory.
We derive necessary and sufficient conditions for a 
family of graph states to be in TQO-1,
which is a class of  quantum error correction
code states whose code distance scales macroscopically with the number of 
physical qubits. Using these criteria, we consider a number of specific graph 
families, including the star and complete graphs, and the line graphs of 
complete and completely bipartite graphs, and discuss which are topologically 
ordered and how to construct the codewords. The formalism is then employed to 
construct several codes with macroscopic distance, including a three-dimensional 
topological code generated by local stabilizers that also has a macroscopic 
number of encoded logical qubits. The results indicate that graph states provide 
a fruitful approach to the construction and characterization of topological 
stabilizer quantum error correction codes.

\end{abstract}
\maketitle

\section{Introduction}
Topologically ordered states have been the focus of much research activity 
because of their novel features and properties in condensed matter 
physics~\cite{Xiao2011,Wang2017,Rachel2018} and their potential use in 
fault-tolerant quantum information processing~\cite{KITAEV2003Fault,
Mong2014Universal,Kraus2013Braiding,Sau2010Generic,Alicea2011NonAbelian}. 
The characterization and detection of topological order is a non-trivial task 
in general because of intrinsically long-range entanglement and the 
difficulty of distinguishing different topological phases using only local 
operators~\cite{Chen2010Local}. Approaches to the characterization of 
topological order include the entanglement 
entropy\cite{Kitaev2006Topological,Levin2006Detecting}, quantum Fisher 
information~\cite{Zhang2018Characterization}, quantum circuit complexity for 
state generation~\cite{Chen2010Local}, and irreducible multiple-party 
correlation~\cite{Zhou2008Irreducible,Liu2016Irreducible}, among others. These
suffer from a number of problems. For example, the determination of 
topological entanglement entropy relies heavily on the smoothness of the 
boundary; otherwise, subtle correction terms may be 
needed~\cite{Papanikolaou2007,Williamson2019}. State generation complexity also
does not provide sufficient information about the topological order; 
topologically ordered states cannot be prepared within the quantum cicuit model
in constant depth~\cite{liao2021graph}, and there also exist states that cannot
be generated in constant depth but are nevertheless not topologically ordered, 
such as GHZ states and cube states~\cite{Aharonov2018Quantum}.

There is an important bilateral relationship between quantum error correction 
codes (QECCs)~\cite{QEC2013} and topological order. An $[[n,k,d]]$ QECC, where 
$n$ is the number of qubits, $k$ is the number of logical (encoded) qubits, and 
$d$ is the code distance, is defined as a subspace of dimension $2^k$ 
($1\leq k\leq n$) in Hilbert space $\mathscr{H}_2^{\otimes n}$ that is able 
correct arbitrary independent errors that occur on $t=\lfloor(d-1)/2\rfloor$ 
qubits. On the one hand, QECCs can be constructed from topologically ordered 
states, such as the toric code~\cite{KITAEV2003Fault}, color 
code~\cite{Bombin2006Topological}, and surface 
code~\cite{Dennis2002topological}. On the other hand, the formulation 
of QECCs can itself provide a natural characterization of topologically ordered 
states: QECC states with `macroscopic' distance~\cite{Bravyi2010Topological}, i.e.\ {where the code distance $d$ scales as $d\sim n^c$ where $0<c\le1$. The quantum states in the macroscopic-distance QECCs are called TQO-1 states.}

Graph theory has been used extensively to construct 
QECCs~\cite{Schlingemann2001quantum,tonchev2002error,MacKay2004sparse,
danielsen2005self,couvreur2013construction,cafaro2014}, providing a rich 
toolkit for the exploration of 
topological codes based on graph states. Furthermore, restrictions on the 
geometrical locality of stabilizer generators or the dimension of 
physical qubits pose fundamental limits on the performance of stabilizer 
QECCs~\cite{Bravyi2010Tradeoffs}; as there are no such restrictions on the 
graphs associated with stabilizer codes, one can potentially circumvent these
limits and construct better LDPC codes.
In recent work~\cite{liao2021graph}, the toric code 
stabilizer states were mapped to graph states using local Clifford 
operations~\cite{Van2004Graphical}, giving a specific graph configuration 
corresponding to topologically ordered states and opening a new window into the 
construction of topological codes. The graph associated with the toric code was 
found to be composed of only two simple graphs (star and half graphs). This 
naturally leads to the possibility that particular graphs are most conducive to 
the construction of topological QECCs, and for the characterization of 
topological order. In this work, we show how to decide if a given family of 
graphs corresponds to TQO-1 states.

In addition to providing  a characterization of topological order, our 
graph-theoretical framework can be used to construct QECCs with macroscopic 
distance and macroscopic scaling in the number of encoded qubits. This behavior
is expected to be important for fault tolerant logical operations on practical
quantum devices, and might also shed light on an quantum complexity. For 
example, the existence of good quantum low-density parity check (LDPC) 
codes~\cite{Kovalev2013,Breuckmann2021}, where both $d$ and $k$ scale linearly with~$n$ and the 
weight of the stabilizer generators is constant, is related to {important quantum complexity problems}~\cite{aharonov2013quantum,Eldar2017local,freedman2013quantum}, so 
the exploration of topological codes might shed light on this open quantum 
complexity problem. Most generally, the construction of new topological codes 
provides insights about the nature of the special order underpinning 
topological states. 

In this work, we derive necessary and sufficient conditions for graph states to 
be topologically ordered, under the definition of TQO-1. To illustrate the use of
this condition, we first discuss several families of graph states that are not
in TQO-1: graphs with constant vertex degree, and the star and complete graphs
whose maximum vertex degree grows with the number of vertices. We then show
that, in contrast, the family of graph states corresponding to multiple copies 
of the star graph is in TQO-1, as are the graphs associated with the toric 
code (hitherto referred to as the toric graph), and the line graphs of the 
complete graph and the complete bipartite graph; the last three all correspond 
to $[[n,1,\sqrt{n}]]$ QECCs. By embedding a classical LDPC code into the 
multi-star graph state, we can also obtain {a $[[ckd,k,d]]$ QECC where $c$ is an independent constant.} These 
families of states all share a common feature, which is necessary (but not 
sufficient) for the graphs to be in TQO-1: the maximum vertex degree must scale 
macroscopically. Using these insights, we construct a topological QECC based on 
a generalization of the toric graph. This QECC corresponds to qubits located on 
the vertices of a three-torus, stabilized by six-local Pauli operators, and 
corresponds to a $[[n,n^{1/3},n^{1/3}]]$ QECC. The distance and the number of 
logical qubits of our constructed code is similar to some of the known 3D 
topological codes~\cite{Vijay2016,Bravyi2011,Shirley2019}.

This work is organized as follows. The background information is reviewed in 
Sec.~\ref{sec:background}. 
The criteria for a family of graph states to be in 
TQO-1 are derived in Sec.~\ref{sec:topological_graph_state}. 
In Sec.~\ref{sec:ExampleAnalysis}, several families of graph states are 
discussed that both do and do not satisfy the criteria for TQO-1. 
For these graph state that are TQO-1, QECCs with macroscopic distance are also 
constructed from the graph states. Sec.~\ref{sec:3DToircGraphCode} provides an 
example of a topological QECC derived from an extension of the toric graph,
with 
qubits on the vertices of three-torus. The results are discussed in 
Sec.~\ref{sec;discussion}. Some technical details are included in the 
appendices.

\section{Background}
\label{sec:background}
In this section, the background formalism to the main results is provided. 
Key definitions and concepts include the TQO-1 class, the quantum error 
correction conditions, graph theory, and graph states. Some of the definitions 
are rephrased so that they can be repurposed for this work.

\subsection{TQO-1 class and QECC condition}
This subsection provides a review of the class of quantum states called 
TQO-1~\cite{Bravyi2010Topological}. This class is defined using QECC and 
captures some of the main features of topological order. As TQO-1 is described 
using QECC, the quantum error correction conditions are also discussed. 

The state~$\Ket{\mathbf{x}}$ is the computational basis state labeled by 
$\mathbf{x}\in\set{0,1}^n$ in the $n$-qubit Hilbert space $\mathscr{H}_2^{\otimes n}$.
A quantum state is understood to mean a trace-class positive-definite operator 
on Hilbert space, and a pure state is any idempotent operator in this class.
Each pure state is identified with a specific vector in the Hilbert space. We
define $\mathbf{P}_{*}(\mathbb{C}):=\bigoplus_{n\in\mathbb{N}}
\mathbf{P}_{2^n}(\mathbb{C})$, the direct sum of all complex projective spaces 
of dimension that is a power of 2; here, $\mathbb{N}_s\subseteq\mathbb{N}$ is a subset of integers containing countably infinite elements.

\begin{definition}[Quantum-state oracle]
A quantum-state oracle
\begin{equation}
\textsc{qState}:\mathbb{N}_s\to \mathbf{P}_{*}(\mathbb{C}):
n\mapsto\psi_n
:=\begin{pmatrix}
\alpha_0\\ \vdots\\ \alpha_{2^n-1}
\end{pmatrix}
\end{equation}
maps a non-negative integer to a vector in a complex projective space. 

\end{definition}
\begin{definition}[Family of quantum states]
Given \textsc{qState}, a family of quantum states for this oracle is $\mathtt{S}:=\Set{\Ket{\psi_n} \in\mathscr{H}_2^{\otimes n} }_{n\in\mathbb{N}_s} $,
where 
\begin{equation}
\label{eq:qstate_psi^n}
\Ket{\psi_n} 
:= \sum_{\mathbf{x}\in\set{0,1}^n} \alpha_\mathbf{x} \Ket{\mathbf{x}}.
\end{equation}
\end{definition}
\noindent Whenever the term `a family of quantum states' is used in this work, 
a \textsc{qState} oracle and $\mathbb{N}_s$ are implied and are often not 
stated explicitly. For $N\in\mathbb{N}$, 
$\left(\mathbf{P}_{N}(\mathbb{C})\right) ^{\otimes *}:=\bigoplus_{i\in[N]} 
\left(\mathbf{P}_{N}(\mathbb{C})\right)^{\otimes i}$ is the direct of sum of 
$i$-fold ($i\in[N]:=\set{1,\ldots,N}$) $N$-dimensional complex projective 
spaces, and $\left(\mathbf{P}_{*}(\mathbb{C})\right) ^{\otimes *}
:=\bigoplus_n \left(\mathbf{P}_{2^n}(\mathbb{C})\right) ^{\otimes *}$.
\begin{definition}
A quantum-code oracle maps two non-negative integers~$n$ and $k$ to $2^k$ 
independent vectors in the complex projective space 
$\mathbf{P}_{2^n}(\mathbb{C})$:
\begin{align}
\textsc{qCode}&: \mathbb{N}\times \mathbb{N} \to \left(\mathbf{P}_{*}(\mathbb{C})\right) ^{\otimes *}\\
&:(n,k) \mapsto (\psi_n^{(0)},\ldots,\psi_n^{(2^k-1)}).\nonumber
\end{align}
\end{definition}

\begin{definition}[Family of subspaces]
Given a \textsc{qCode} oracle and function $k:\mathbb{N\to\mathbb{N}}$, a family of subspaces is
\begin{equation}
\label{eq:familysubspaces}
\mathtt{C}=\Set{\mathscr{C}_n:=\operatorname{span}\Set{\Ket{\psi_n^{(0)}},\ldots,\Ket{\psi_n^{(2^{k_n}-1)}}} }_{n\in\mathbb{N}_s},
\end{equation}
where $\psi_n^{(i)}$ is one of the vectors in the range of $\textsc{qcode}(n,k_n)$ and $\Ket{\psi_n^{(i)}} \in \mathscr{H}_2^{\otimes n} (0\le i\le 2^{k_n}-1)$ is constructed according to Eq.~(\ref{eq:qstate_psi^n}).
\end{definition}

\begin{definition}[Family of QECCs]
Given a  family of subspaces~$\mathtt{C}=\Set{\mathscr{C}_n}_{n\in\mathbb{N}_s}$
and $d:\mathbb{N}\to\mathbb{N}$, if every $\mathscr{C}_n\in\mathtt{C}$ is also a QECC of distance $d_n$,
then $\mathtt{C}$ is a family of $[[n,k,d]]$ QECCs.
\end{definition}
\noindent In this work, whenever the `family of subspaces / codes' term is 
employed, functions $k$ and $d$ are implied yet not always stated explicitly.
A family of $[[n,k,d]]$ QECCs has `macroscopic distance' if  
$d\in\operatorname{sublin}(n):=\cup_{0<c\le1}\Theta(n^{c})$, which is the set 
of polynomials with power less than 1.

\begin{definition}[Class TQO-1]
TQO-1 is the {set} of families of quantum states such that for every family 
$\mathtt{S}=\Set{\Ket{\psi_n} }_{n\in\mathbb{N}_s} \in \text{TQO-1}$,
there exists a family of macroscopic-distance QECCs $\mathtt{C}=\Set{\mathscr{C}_n}_{n\in\mathbb{N}_s}$,
satisfying $\Ket{\psi_n}\in \mathscr{C}_n \forall n\in\mathbb{N}_s$. 
\end{definition}

In order to evaluate the distance of a QECC, we restate the quantum error 
correction condition from~\cite{Knill1997Theory,nielsen2000quantum}. Let 
$\mathcal{D}(\mathscr{H}_2^{\otimes n})$ denote the set of density matrices on 
$n$ qubits.
\begin{theorem}
[Quantum error correction condition~\cite{Knill1997Theory,nielsen2000quantum}]
\label{theorem:qerror_corr_condition}
Suppose $\mathscr{C}\subset \mathscr{H}_2^{\otimes n}$ is a QECC,  $P$ is the projector into $\mathscr{C}$ from $\mathscr{H}_2^{\otimes n}$, 
$\mathcal{E}:\mathcal{D}(\mathscr{H}_2^{\otimes n})\to \mathcal{D}(\mathscr{H}_2^{\otimes n})$ is a quantum channel with Kraus operators $\left\{E_{i}\right\}_{i\in[2^n]}$.
A necessary and sufficient condition for the existence of an error-correction operation $\mathcal{R}$ correcting $\mathcal{E}$ on $\mathscr{C}$ is that
\begin{equation}
P E_{i}^{\dagger} E_{j} P=\alpha_{i j} P  
\end{equation}
for some Hermitian matrix $\alpha \in\mathbb{C}^{2^n\times 2^n}$.
\end{theorem}
A simplified quantum error correction condition derived from the 
Theorem~\ref{theorem:qerror_corr_condition} is used later in this work.
The 
Pauli group on~$n$ qubits is
$\mathcal{P}_n := \set{\pm i,\pm1}\times\set{I,X,Y,Z}^{\otimes n}$, where 
$X,Y,Z$ are 1-qubit Pauli operators. The weight of a Pauli operator $\text{wt}(O)$ 
corresponds to the number of qubits acted on by $O$ non-trivially. Then,
$\mathcal{P}_n^t:=\{O\in \mathcal{P}_n| w(O)\le t\}$ contains every operator in 
the Pauli group whose weight is no greater than $t$.
\begin{corollary}
\label{coro:qcode_distance_d}
Given $d<n\in\mathbb{N}$ and two states $\Ket{\phi},\Ket{\psi}\in \mathscr{H}_2^{\otimes n}$,
then $\operatorname{span}_{\mathbb{C}}\{\Ket{\phi},\Ket{\psi}\}  \subseteq \mathscr{H}_2^{\otimes n}$ is a $[[n,1,d ]]$ QECC iff
\begin{align}
\label{eq:phiOphi=psiOpsi}
\braket{\phi|O|\phi}= &\braket{\psi|O|\psi}\\
\label{eq:phiOpsi=0}
\braket{\phi|O|\psi}=&0
\end{align}
hold $\forall O\in\mathcal{P}_n^{d-1}$.
\end{corollary}
\begin{proof}
A quantum error code of distance $d$ can correct any 
errors on less than or equal to $t=\frac{d-1}{2}$ qubits. Thus, to prove $C$ is 
a distance-$d$ code is equivalent to proving that $\mathcal{P}_n^t$ is a set of 
correctable errors on $C = \operatorname{span}_{\mathbb{C}}\{|\psi_1\rangle,|\psi_2\rangle \}$, as any 
Hermitian operator on $t$ qubits can be decomposed into the sum of operators in 
$\mathcal{P}_n^t$.

For any $E_i,E_j \in \mathcal{P}_n^t $, one has 
$E_j^\dagger E_i \in \mathcal{P}_n^{2t}$, so that
\begin{equation}
P = |\psi_1\rangle\langle \psi_1|+ |\psi_2\rangle\langle \psi_2|    
\end{equation}
and
\begin{eqnarray}
PE_j^\dagger E_i P&=&|\psi_1\rangle\langle \psi_1|
\langle \psi_1|E_j^\dagger E_i|\psi_1\rangle \nonumber \\
&+&|\psi_2\rangle\langle \psi_2|
\langle \psi_2|E_j^\dagger E_i|\psi_2\rangle \nonumber \\
&+&|\psi_1\rangle\langle \psi_2|
\langle \psi_1|E_j^\dagger E_i|\psi_2\rangle \nonumber \\
&+&|\psi_2\rangle\langle \psi_1|
\langle \psi_2|E_j^\dagger E_i|\psi_1\rangle.
\end{eqnarray}
Thus, $PE_j^\dagger E_iP = \alpha_{ij}P$ is equivalent to
\begin{align}
\langle\psi_1|E_j^\dagger E_i|\psi_1\rangle =&\langle\psi_2|E_j^\dagger E_i|\psi_2\rangle = \alpha_{ij}\\
\langle\psi_1|E_j^\dagger E_i|\psi_2\rangle =& 0 .
\end{align}
The requirement that $\alpha$ is a Hermitian matrix
\begin{equation}
\alpha_{ij}^* = \braket{\psi_1|E_j^\dagger E_i|\psi_1}^* = \braket{\psi_1|E_i^\dagger E_j|\psi_1} = \alpha_{ji}
\end{equation}
is also satisfied.
\end{proof}

\subsection{Graph theory}
\label{sec:graph_theory}
Graph theory has been used extensively to construct QECCs~\cite{Schlingemann2001quantum,
tonchev2002error,MacKay2004sparse,danielsen2005self,
couvreur2013construction}, providing a rich toolkit for the exploration of 
topological codes based on graph states. 
A graph $G=(V,E)$ is composed of a set of vertices $V$ and a set of edges $E$, 
in which an edge $e=(v_i,v_j)$ corresponds to a pair of vertices. In this work,
we only consider undirected graphs, where $(v_i,v_j)$ and $(v_j,v_i)$ define
the same edge. The graph~$G$ can be represented by its adjacency matrix 
$A\in\mathbb{Z}_2^{|V|\times|V|}$ such that
\begin{equation}
A_{ij}=1 \iff  (v_i,v_j)\in E.  
\end{equation}

One can obtain a subgraph of $G$ by deleting an edge $e$, which is denoted by
$G\backslash e$, or deleting a vertex $v$ and every edge incident to this 
vertex, which is denoted by $G-v$. In this work, only subgraphs obtained by 
edge deletion are considered. The degree of a vertex $\operatorname{deg}(v)$ is 
the number of edges which are incident to $v$. The handshaking lemma states
that for every graph, the number of vertices with odd degree is 
even~\cite{biggs1986}.

In this work, we consider a small number of special graphs. The complete graph, 
denoted by $K_n$, has every pair of vertices connected by an edge, so that
there are ${n(n-1)}/{2}$ edges. The graph $G$ is bipartite if its vertices can 
be partitioned into two complementary subsets $X\sqcup Y=V$, such that there is 
no edge in $G$ connecting two vertices in $X$ or two vertices in $Y$. The graph
$G$ is a complete bipartite graph if all vertices in $X$ are connected to all 
vertices in $Y$; this is usually denoted as $K_{n,m}$ when $|X|=n$ and $|Y|=m$.
The line graph of $G$, denoted by $L(G)=(V^\prime,E^\prime)$, is constructed as 
follows. Every edge in $E$ is mapped to a vertex in $V^\prime$, and two 
vertices $v_1^\prime$ and $v_2^\prime$ in $V^\prime$ are connected iff their 
corresponding edges $e_1$ and $e_2$ in $E$ share a common vertex. The line 
graph of the complete graph $L(K_m)$ is therefore an ${m(m-1)}/{2}$-vertex 
graph, and the line graph of the complete bipartite graph $L(K_{m,m})$ has 
$m^2$ vertices.

A path in a graph $G$ is sequence of distinct edges 
$\{E_1,\ldots,E_i,\ldots\}$, in which every edge $E_i$ (except for the first 
one) starts with the vertex that edge $E_{i-1}$ ends with. A cycle is a path 
where the last edge ends at the vertex that the first edge started with. An 
Eulerian cycle in a finite graph is the cycle that visits each edge exactly 
once. Based on Euler's theorem, a connected graph has an Eulerian cycle iff 
every vertex has even degree~\cite{biggs1986}.

As bitstrings are extensively employed in this work, it behooves us to define 
some basic notation related to bitstring operations. For two bitstrings 
$\mathbf{k},\mathbf{l}\in\set{0,1}^n$, $(\mathbf{k}+\mathbf{l})\in\set{0,1}^n$ is the bitwise XOR; 
$(\mathbf{k}\vee \mathbf{l})\in\set{0,1}^n$ is the bitwise OR; and 
$\mathbf{k}\cdot \mathbf{l} = \sum_{i\in[n]} \mathbf{k}_i\cdot \mathbf{l}_i \mod 2 \in\set{0,1}$ is the `inner 
product' between two bitstrings.
Given matrix $A\in\mathbb{Z}_2^{n\times n}$ and bitstring $\mathbf{k}\in\set{0,1}^n$, 
$(A\cdot \mathbf{k})\in\set{0,1}^n$ is the bitstring mapped from $\mathbf{k}$ by $A$ such that
$(A\cdot \mathbf{k})_i = \sum_{j\in[n]} A_{ij}\mathbf{k}_j \mod 2$.  A set of bitstrings 
$B=\Set{\mathbf{k}^1,\ldots,\mathbf{k}^m}\subseteq \mathbb{Z}_2^{n}$ is independent if for every 
subset $B_s\subseteq B$ the following holds:
\begin{equation}
\sum_{\mathbf{k}\in B_s} \mathbf{k} \ne 0^{n}.   
\end{equation}
The $i$-th basis bitstring is denoted by $\mathbf{b}^i$, $(i\in[n])\in\set{0,1}^n$, 
i.e.\ $\mathbf{b}^i_j=1$ iff $j=i$.

For a graph $G=(V,E)$, $\mathcal{E}(G)$ denotes the set of all subsets of $E$.
It is easy to see for arbitrary two subsets $E_1,E_2\subseteq E$, 
$E_1\triangle E_2$ is also a subset of $E$, where $\triangle$ is the set 
symmetric difference. As a result, $(\mathcal{E}(G), \triangle)$ spans a vector 
space of dimension $|E|$ over $\mathbb{Z}_2$. Alternatively, one can assign a 0 
or 1 to each edge and obtain a bitstring $\mathbf{k}\in \set{0,1}^{|E|}$, the set of 
which also spans a vector space. There is therefore a bijective mapping between 
the bitstring and edge subsets:
\begin{equation}
\mathbf{k}\in\set{0,1}^{|E|} \leftrightarrow E_\mathbf{k}:=\Set{e\in E| \mathbf{k}_e=1}.
\end{equation}
The symmetric difference and bitwise XOR are mapped to one another, and the
vector space $(\mathcal{E}(G),\triangle)$ is isomorphic to 
$(\mathbb{Z}_2^{|E|},+)$. 

Let $G_\mathbf{k}=(V_\mathbf{k},E_\mathbf{k})$ denote the subgraph of $G$ comprised of edges corresponding 
to the non-zero entries in $\mathbf{k}$. Then, two bitstrings $\mathbf{k}^i$ and $\mathbf{k}^j$ are 
orthogonal, $\mathbf{k}^i\cdot \mathbf{k}^j=0$, iff the subgraph $G_{\mathbf{k}^i}$ and $G_{\mathbf{k}^j}$ share an 
even number of edges. Consider for example the complete graph $K_4$ with edge
set $\{e_1,e_2,e_3,e_4,e_5,e_6\}$. Bitstrings $\mathbf{k}^1=111100$ and $\mathbf{k}^2=100110$
correspond to the edge subsets $\{e_1,e_2,e_3,e_4\}$ and $\{e_1,e_4,e_5\}$,
respectively. The bitstring $\mathbf{k}^3=\mathbf{k}^1+\mathbf{k}^2=011010$ corresponds to edge subset 
$\{e_2,e_3,e_5\}$, which equals $\{e_1,e_2,e_3,e_4\}\triangle\{e_1,e_4,e_5\}$.
$G_{\mathbf{k}^1}$ and $G_{\mathbf{k}^2}$ have two common edges and $\mathbf{k}^1\cdot \mathbf{k}^2=1+1 \mod 2=0$;
$G_{\mathbf{k}^2}$ and $G_{\mathbf{k}^3}$ share only one edge and $\mathbf{k}^2\cdot \mathbf{k}^3=1$.

\subsection{Graph states and graph basis states}
Operators in the Pauli group $\mathcal{P}_n$ without the prefactor 
$\set{\pm i,\pm1}$ can be written as 
\begin{equation}
O=X^\mathbf{k}Z^\mathbf{l}:= \bigotimes_{i\in[n]} X_i^{\mathbf{k}_i}\bigotimes_{j\in[n]} Z_j^{\mathbf{l}_j}, 
\end{equation}
where $\mathbf{k},\mathbf{l}\in\set{0,1}^n$ and $X_i^0=Z_j^0=I$. 
The set ${\mathcal S}:=\{S\in{\mathcal P}_n |\; S\Ket{\psi} = \Ket{\psi}\}$ is 
said to stabilize the state 
$\ket{\psi}\in\mathscr{H}_2^{\otimes n}$~\cite{gottesman1997}. 
The set of states simultaneously stabilized by $m$ independent operators 
$\{S_1,\ldots,S_m\}$ from $\mathcal{P}_n$ then yields a state subspace 
$V_\mathcal{S}\subseteq \mathscr{H}_2^{\otimes n}$ of dimension $2^{n-m}$. 
A stabilized subspace $V_\mathcal{S}$ that corresponds to a $[[n,n-m,d]]$ QECC
is denoted a $[[n,n-m,d]]$ stabilizer QECC. When $m=n$, the subspace contains 
only one state called the stabilizer state, and the~$n$ independent operators 
are the generators of ${\mathcal S}$.

Graph states are special stabilizer states where the stabilizer generators are 
related to simple graphs~\cite{Hein2004Multiparty}. Given a graph $G=(V,E)$, 
where $|V|=n$, the corresponding graph state is
\begin{align}
\label{eq:graph_state_CZij}
    \Ket{G} :=  \prod_{(i,j)\in E} \operatorname{CZ}(i,j) H^{\otimes n} 
\Ket{0^n}
\in\mathscr{H}_2^{\otimes n},
\end{align}
in which $i,j\in [n]$ labels vertices (qubits) in graph $G$ and 
$\operatorname{CZ}(i,j)$ is the controlled-$Z$ gate. The stabilizer generators 
for state~(\ref{eq:graph_state_CZij}) are
\begin{equation}
\label{eq:grpah_state_S_i}
\{S_i\}_{i\in[n]} = \left\{X_{i}  \prod_{(i,j)\in E } Z_{j}\right\}.
\end{equation}
For a given graph state~$\Ket{G}$, the set
\begin{equation}
\label{eq:graph_basis}
\set{\Ket{\mathbf{h}}_G:=\otimes_iZ_i^\mathbf{h}\Ket{G}=Z^\mathbf{h}\Ket{G};
\mathbf{h}\in\{0,1\}^n}
\end{equation}
is an orthogonal basis in $\mathscr{H}_2^{\otimes n}$, where state~$\Ket{\mathbf{h}}_G$ 
is called a graph basis state~\cite{hein2006entanglement}. Evidently, the graph 
state~$\Ket{G}$ is~$\Ket{0^n}_G$. The subscript $G$ is used in the graph basis 
state notation in order to prevent any confusion with the computational basis. 

\section{Topological graph state}
\label{sec:topological_graph_state}
This section presents necessary and sufficient conditions for a family of graph 
states to be in TQO-1
and begins with a definition and notation.
A family of graph states is a family of states with the restriction that all 
states are graph states. Given a family of graph states 
$\mathtt{S}=\{\Ket{G_n}\in\mathscr{H}_2^{\otimes n}\}_{n\in\mathbb{N}_s}$, is
$\mathtt{S}$ in the TQO-1 class, defined using a macroscopic-distance QECC 
containing at least 2 quantum states? This question can be answered by 
investigating if there exists another family of states 
$\Set{\Ket{\psi_n}}_{n\in\mathbb{N}_s}$ such that 
$\left\{\mathscr{C}_n=\operatorname{span}_{\mathbb{C}}\{\Ket{G_n},
\Ket{ \psi_n\}} \right\}_{n\in\mathbb{N}_s}$ is a family of QECCs with 
macroscopic distance.

It is convenient and insightful to start with a special case: given 
$n,d\in\mathbb{N}$ satisfying $d<n$, and a graph state 
$\Ket{G}\in\mathscr{H}_2^{\otimes n}$, decide the existence of a state 
$\Ket{\psi}\in\mathscr{H}_2^{\otimes n}$ such that 
$\operatorname{span}_{\mathbb{C}}\{\Ket{G},\Ket{\psi}\}$ is a $[[n,1,d]]$ QECC.
In principle, every state orthogonal to~$\Ket{G}$ could be a potential 
candidate~$\Ket{\psi}$ to make a $d$-distance QECC. In order to simplify the 
analysis, one can restrict to the case where the QECC 
$\operatorname{span}_{\mathbb{C}}\{\Ket{G},\Ket{\psi}\}$ is a stabilizer QECC. 
This restriction leads to the following result, the proof for which is given in 
the Appendix~\ref{appdix:stabilized_subspace}
\begin{lemma}
\label{lemma:stabilizer_qecc_graph_basis}
Given $n\in\mathbb{N}$ and two orthogonal quantum states 
$\Ket{G},\Ket{\psi}\in\mathscr{H}_2^{\otimes n}$, in which~$\Ket{G}$ is a graph 
state, the subspace $\operatorname{span}_{\mathbb{C}}\{\Ket{G},\Ket{\psi}\}
\subseteq \mathscr{H}_2^{\otimes n} $ is a stabilized subspace iff~$\Ket{\psi}$ 
is a graph basis state~$\Ket{\mathbf{h}}_G$ for some $\mathbf{h}\in\set{0,1}^n$.
\end{lemma}
 
It therefore suffices to analyze the case where~$\Ket{\psi}$ is a graph basis 
state~$\Ket{\mathbf{h}}_G$. Given~$\Ket{G}$ and~$\Ket{\mathbf{h}}_G$, the next task is to 
determine if $\operatorname{span}\{\Ket{G},\Ket{\mathbf{h}}_G\}$ is a $[[n,1,d]]$ QECC,
which can be accomplished using Corollary~\ref{coro:qcode_distance_d}. 
Alternatively, one must determine if conditions 
$\braket{G|O|G}={}_G\!\braket{\mathbf{h}|O|\mathbf{h}}_G$ and $\braket{G|O|\mathbf{h}}_G=0$ are satisfied 
for all $O\in\mathcal{P}_n^{d-1}$. As any operator in the Pauli group 
$\mathcal{P}_n$ can be expressed as $O=X^\mathbf{k} Z^\mathbf{l}$, the condition 
$O=X^\mathbf{k} Z^\mathbf{l}\in\mathcal{P}_n^{d-1}$ in Corollary~\ref{coro:qcode_distance_d} is 
equivalent to 
\begin{align}
(\mathbf{k},\mathbf{l})\in \mathcal{B}_n^{d-1}:=\{(\mathbf{m}_1,\mathbf{m}_2)\in\{0,1\}^n\times \{0,1\}^n|\\
\text{wt}(\mathbf{m}_1\vee \mathbf{m}_2)\le d-1\},    \nonumber
\end{align}
in which ${\rm wt}$ is the Hamming weight. 

\begin{lemma}
\label{lemma:iff_h_stabilzier_qecc}
Given $d,n\in\mathbb{N}$ satisfying $d\le n$, the graph state 
$\Ket{G}\in\mathscr{H}_2^{\otimes n}$, and $\mathbf{h}\in\set{0,1}^n\backslash\{0^n\}$, 
$\operatorname{span}_{\mathbb{C}}\{\Ket{G},\Ket{\mathbf{h}}_G\}$ is an $[[n,1,d]]$ 
stabilizer QECC iff
\begin{align}
\label{eq:h-kl=0-kl}
{}_G\!\braket{\mathbf{h}|X^\mathbf{k}Z^\mathbf{l}|\mathbf{h}}_G= &\braket{0^n|X^\mathbf{k}Z^\mathbf{l}|0^n}_G, \forall (\mathbf{k},\mathbf{l})\in\mathcal{B}_n^{d-1},\\
\label{eq:hkl0=0}
{}_G\!\braket{\mathbf{h}|X^\mathbf{k}Z^\mathbf{l}|0^n}_G=&0, \forall (\mathbf{k},\mathbf{l})\in\mathcal{B}_n^{d-1},
\end{align}
holds.
\end{lemma}

In order to proceed further, one must calculate the value of 
${}_G\!\braket{\mathbf{h}|X^\mathbf{k}Z^\mathbf{l}|\mathbf{g}}_G$ for given $\mathbf{h},\mathbf{g}\in\set{0,1}^n$.
The following result is proven in Appendix~\ref{appendix:graphbasisexpec}.
\begin{lemma}
\label{lemma:graphbasisexpec}
Given $n$-vertex graph $G$ and $\mathbf{h},\mathbf{g},\mathbf{k},\mathbf{l}\in\set{0,1}^n$, $A\in\mathbb{Z}_2^{n\times n}$  is the adjacency matrix  of $G$.
The value of ${}_G\!\braket{\mathbf{h}|X^\mathbf{k}Z^\mathbf{l}|\mathbf{g}}_G$ is
\begin{align}
{}_G\!\braket{\mathbf{h}|X^\mathbf{k}Z^\mathbf{l}|\mathbf{g}}_G =
\begin{cases}
(-1)^{\mathbf{h}\cdot \mathbf{k}+\sigma(A,\mathbf{k})},& \text{if}~A\cdot \mathbf{k}+\mathbf{l}=\mathbf{h} +\mathbf{g}, \\
0,&\text{otherwise},
\end{cases}
\end{align}
in which
\begin{equation}
\sigma(A,\mathbf{k}) = \sum_{j=2}^n \mathbf{k}_j\left(\sum_{i=1}^{j-1} \mathbf{k}_i A_{ij} \right).
\end{equation}
\end{lemma}

With Lemma~\ref{lemma:graphbasisexpec}, one can easily verify if conditions~(\ref{eq:h-kl=0-kl}) and~(\ref{eq:hkl0=0}) hold. It is convenient at this stage
to introduce subsets of bitstrings which are closely related to these 
conditions. Given $d<n\in\mathbb{N}$, an $n$-vertex graph $G$, and its 
adjacency matrix $A\in\mathbb{Z}_2^{n\times n}$, four sets of length-$n$ 
bitstrings can be defined as follows: 
\begin{align}
\label{eq:Z(G,n,d)}
Z(G,n,d):=& \left\{\mathbf{k}\in\set{0,1}^n|(\mathbf{k},A\cdot \mathbf{k})\in\mathcal{B}_{n}^{d-1}
\right\},\\
Z^\perp(G,n,d):=&\left\{\mathbf{k}\in\set{0,1}^n| \right.
 \mathbf{k}\cdot \mathbf{l}= 0\forall l\in Z(G,n,d)\},\\
W(G,n,d):=&\{\mathbf{k}\in\{0,1\}^n|\nonumber \\
&\quad \mathbf{k}=A\cdot m+\mathbf{l}, (m,\mathbf{l})\in\mathcal{B}_n^{d-1}\},\\
C(G,n,d):=&Z^\perp(G,n,d)\backslash W(G,n,d).
\end{align}
When focusing on specific graphs, $G$ and~$n$ are implicitly known, so the
notation is sometimes simplified as $Z(d)\equiv Z(G,n,d)$ (likewise for the
other three sets) when it does not cause any confusion. With these definitions, 
the quantum error correction condition for 
$\operatorname{span}_{\mathbb{C}}\{\Ket{G},\Ket{\mathbf{h}}_G\}$ can be expressed 
compactly as follows.
\begin{lemma}
\label{lemma:CGnd_qecc}
Given $d,n\in\mathbb{N}$ satisfying $d\le n$, the graph state 
$\Ket{G}\in\mathscr{H}_2^{\otimes n}$, and $\mathbf{h}\in\set{0,1}^n\backslash\{0^n\}$, 
$\operatorname{span}_{\mathbb{C}}\{\Ket{G},\Ket{\mathbf{h}}_G\}$ is a $[[n,1,d]]$ QECC 
iff $\mathbf{h}\in C(G,n,d)$.
\end{lemma}
\begin{proof}
It suffices to prove that Eqs.~(\ref{eq:h-kl=0-kl}) and~(\ref{eq:hkl0=0}) 
are equivalent to $\mathbf{h}\in C(G,n,d)$. Consider Eq.~(\ref{eq:h-kl=0-kl}). If 
$A\cdot \mathbf{k}\ne \mathbf{l}$, then ${}_G\!\braket{\mathbf{h}|X^\mathbf{k}Z^\mathbf{l}|\mathbf{h}}_G=\braket{0^n|X^\mathbf{k}Z^\mathbf{l}|0^n}_G=0$;
if $A\cdot \mathbf{k}=\mathbf{l}$, then ${}_G\!\braket{\mathbf{h}|X^\mathbf{k}Z^\mathbf{l}|\mathbf{h}}_G=(-1)^{\mathbf{h}\cdot \mathbf{k}}$ and 
$\braket{0^n|X^\mathbf{k}Z^\mathbf{l}|0^n}_G=0$. Therefore, Eq.~(\ref{eq:h-kl=0-kl}) is 
equivalent to $\mathbf{h}\cdot \mathbf{k}=0 $ when $A\cdot \mathbf{k}=\mathbf{l}$ and $(\mathbf{k},\mathbf{l})\in\mathcal{B}_n^{d-1}$,
which can be compactly described as $\mathbf{h}\in Z^\perp(G,n,d)$.

For Eq.~(\ref{eq:hkl0=0}), one has ${}_G\!\braket{\mathbf{h}|X^\mathbf{k}Z^\mathbf{l}|0^n}_G=0$ iff 
$A\cdot \mathbf{k}+\mathbf{l}\ne h$, so Eq.~(\ref{eq:hkl0=0}) is equivalent to 
$\mathbf{h}\ne A\cdot \mathbf{k}+\mathbf{l}~\forall (\mathbf{k},\mathbf{l})\in\mathcal{B}_n^{d-1}$, i.e.\ 
$\mathbf{h}\notin W(G,n,d)$. As both Eqs.~(\ref{eq:h-kl=0-kl}) and~(\ref{eq:hkl0=0}) 
need to be satisfied, $\mathbf{h}\in Z^\perp(G,n,d)$ and $\mathbf{h}\notin W(G,n,d)$, i.e.\
$\mathbf{h}\in C(G,n,d)$.
\end{proof}

From the proof of Lemma~\ref{lemma:CGnd_qecc}, the meaning of the subsets 
defined above are: $Z^\perp(G,n,d)$ represents the graph basis states that 
satisfy the first constraints of the QECC condition; the complementary set of 
$W(G,n,d)$ represents the graph basis states that satisfy the second constraint 
of the QECC condition; and $C(G,n,d)$ represents the graph basis state such 
that the spanned subspace with the graph state is a $[[n,1,d]]$ QECC. With 
Lemma~\ref{lemma:CGnd_qecc} in hand, it is now possible to provide a necessary 
and sufficient condition for the existence of another state~$\Ket{\psi}$ such 
that $\operatorname{span}\{\Ket{G},\Ket{\psi}\}$ is a $[[n,1,d]]$ stabilizer 
QECC.
\begin{lemma}
\label{lemma:SQECC-C(G,n,d)}
Given $d,n\in\mathbb{N}$ satisfying $d\le n$ and the graph state 
$\Ket{G}\in\mathscr{H}_2^{\otimes n}$, 
$\Ket{G}$ is in a $[[n,1,d]]$ stabilizer QECC iff the membership class 
$C(G,n,d)\neq\emptyset$.
\end{lemma}
\begin{proof}
If $C(G,n,d)$ is non-empty, assume $h$ is one of the bitstrings in it. Based on 
Lemma~\ref{lemma:CGnd_qecc}, $\operatorname{span}_{\mathbb{C}}\{\Ket{G},
\Ket{\mathbf{h}}_G\}$ is a $[[n,1,d]]$ stabilizer QECC, so~$\Ket{G}$ is in a 
$[[n,1,d]]$ stabilizer QECC. Or, if~$\Ket{G}$ is in a $[[n,1,d]]$ stabilizer 
QECC, which is assumed to be $\operatorname{span}_{\mathbb{C}}\{\Ket{G},
\Ket{\psi}\}$ for some $n$-qubit state $\Ket{\psi}\in\mathscr{H}_2^{\otimes n}$,
then Lemma~\ref{lemma:stabilizer_qecc_graph_basis} ensures that~$\Ket{\psi}$ is 
a graph basis state~$\Ket{\mathbf{h}}_G$. From Lemma~\ref{lemma:CGnd_qecc}, one obtains
$\mathbf{h}\in C(G,n,d)$, and therefore $C(G,n,d)\ne \emptyset$. 
\end{proof}

The necessary and sufficient condition stated in 
Lemma~\ref{lemma:SQECC-C(G,n,d)} is restricted to the case where 
$\operatorname{span}\{\Ket{G},\Ket{\psi}\}$ is a stabilizer QECC, 
but the defintion of TQO-1 states doesn't require this condition. An immediate
question is if the non-emptiness of $C(G,n,d)$ is still necessary if the 
restriction to a stabilizer QECC is lifted. It turns out that the answer is 
yes: 
\begin{lemma}
\label{lemma:QECC-C(G,n,d)}
Given $d,n\in\mathbb{N}$ satisfying $d\le n$ and graph state $\Ket{G}\in\mathscr{H}_2^{\otimes n}$, 
$\Ket{G}$ is in a $[[n,1,d]]$ QECC iff the membership class 
$C(G,n,d)\neq\emptyset$.
\end{lemma}
\noindent A rigorous proof is provided in Appendix~\ref{appendix:QECC-C(G,n,d)}.

It is useful at this stage to point out that for a given $n$-vertex graph $G$, 
there is a maximum distance $d^{\rm max}$ for all QECCs containing the graph 
state~$\Ket{G}$. Start with the observation that $Z(G,n,d)\subseteq Z(G,n,d+1)$ 
and $W(G,n,d)\subseteq W(G,n,d+1)$; as a result, 
$Z^\perp(G,n,d+1)\subseteq Z^\perp(G,n,d)$ and $C(G,n,d+1)\subseteq C(G,n,d)$.
Next, note that $\mathbf{k}\in Z(G,n,n+1)$ for every $\mathbf{k}\in\set{0,1}^n$. As a result,
$Z^\perp(G,n,n+1)=\emptyset$ and so is $C(G,n,n+1)$. If $d=1$, then the only
bitstring in $Z(G,n,1)$ and $W(G,n,1)$ is $\mathbf{k}=0^n$, so 
$C(G,n,1)=\set{0,1}^n\backslash \set{0^n}$. One therefore has the following 
sequence
\begin{align}
\emptyset=&C(G,n,n+1)\subseteq C(G,n,n)\subseteq C(G,n,n-1)\subseteq 
\cdots \nonumber \\
\subseteq &C(G,n,2)\subseteq C(G,n,1)=\set{0,1}^n \backslash\set{0}.
\end{align}
There exists a critical value $d^{\rm max}\in[n]$ such that 
$C(G,n,d^{\rm max})\ne\emptyset$ while $C(G,n,d^{\rm max}+1)=\emptyset$; in
other words, $d^{\rm max}$ is the largest code distance of all QECCs containing 
the graph state~$\Ket{G}$. If $C(G,n,d)=\emptyset$ for some $d$, then 
$d^{\rm max}< d$; similarly, if $C(G,n,d)\ne \emptyset$ for some $d$, then 
$d^{\rm max}\ge d$. In practice, one can increase the value of $d$ in unit
increments and test the membership of $C(G,n,d)$ until $C(G,n,d^{\rm max}+1)
=\emptyset$.

From Lemma~\ref{lemma:QECC-C(G,n,d)} and the definition of TQO-1, it is now
possible to give a necessary and sufficient condition for a family of graph 
states to be in TQO-1:
\begin{theorem}
\label{theorem:TQO1graphstate}
A family of graph states 
$\mathtt{S}=\{\Ket{G_n}\in\mathscr{H}_2^{\otimes n}\}_{n\in\mathbb{N}_s}$ 
is in TQO-1 iff $d^{\rm max}\in\operatorname{sublin}(n)$.
\end{theorem}
\begin{proof}
If $d^{\rm max}\in\operatorname{sublin}(n)$ such that 
$C(G_n,n,d_n^{\rm max})\neq\emptyset$, assume $\mathbf{h}\in C(G_n,n,d_n^{\rm max})$. 
From Lemma~\ref{lemma:SQECC-C(G,n,d)}, 
$C_n=\operatorname{span}_{\mathbb{C}}\{\Ket{G_n},\Ket{\mathbf{h}}_G\}$ is a
$[[n,1,d_n^{\rm max}]]$ QECC, so $\set{C_n}_{n\in\mathbb{N}_s}$ is a family of 
QECCs with macroscopic distance and contains $\Set{\Ket{G_n}}$. Therefore, 
$\Set{\Ket{G_n}}$ is in TQO-1.

On the other hand, if $\Set{\Ket{G_n}}$ is in TQO-1, then by definition there 
exists $d\in\operatorname{sublin}(n)$ and a family of QECCs $\{\mathscr{C}_n\}$ 
such that every $\mathscr{C}_n$ has distance $d_n$ and contains the graph state 
$\Ket{G_n}$. From Lemma~\ref{lemma:QECC-C(G,n,d)}, $C(G_n,n,d_n)\ne\emptyset$ 
for every $n\in\mathbb{N}_s$. One thus, obtains $d_n^{\rm max} \ge d_n$. From 
the definition of $d^{\rm max}$, $d^{\rm  max }\in\operatorname{sublin}(n)$.
\end{proof}

Let us now generalize Lemma~\ref{lemma:CGnd_qecc} to a $[[n,k,d]]$ QECC, where 
$k>1$, i.e.\ the QECC encodes more than a single logical qubit.
Suppose that the subspace $\mathscr{C} = \operatorname{span}_{\mathbb{C}} 
\left\{\Ket{\psi_1}, \ldots,|\psi_{2^k}\rangle\right\}$ is spanned by $2^k$ 
orthogonal states. Assume that~$\Ket{\psi_1}$ is a graph state and 
$\mathscr{C}$ is a stabilizer QECC. Following the same reasoning behind 
Lemma~\ref{lemma:stabilizer_qecc_graph_basis}, one obtains that every 
$\Ket{\psi_i}$ is a graph basis state~$\Ket{\mathbf{h}^i}_G$, so that 
$\mathscr{C}=\operatorname{span}_{\mathbb{C}} \{\Ket{\mathbf{h}^1}_G,\ldots,
|\mathbf{h}^{2^k}\rangle_G\}$, where $\mathbf{h}^1=0^n$. Using the same argument in the proof of 
Corollary~\ref{coro:qcode_distance_d}, $\mathscr{C}$ is a $[[n,k,d]]$ QECC iff
the conditions
\begin{eqnarray}
{}_G\!\braket{\mathbf{h}^i|O|\mathbf{h}^i}_G &=& {}_G\!\braket{\mathbf{h}^j|O|\mathbf{h}^j}_G;\\
{}_G\!\braket{\mathbf{h}^i|O|\mathbf{h}^j}_G &=&0,
\end{eqnarray}
hold for every $O\in\mathcal{P}_n^{d-1}$ and every $1\le i\ne j\le 2^k$.
Based on the proof of Lemma~\ref{lemma:CGnd_qecc}, the first condition is 
equivalent to $\mathbf{h}^i\in Z^{\perp}(G,n,d)$ for every $i,j\in [2^k]$ and the second 
condition is equivalent to $\mathbf{h}^i+\mathbf{h}^j\notin W(G,n,d)$ for every pair $i\ne j$. 
\begin{corollary}
\label{corollary:QECCgeneralconditionZWset}
Given $d,n\in\mathbb{N}$ satisfying $d\le n$, the graph state 
$\Ket{G}\in\mathscr{H}_2^{\otimes n}$, and $2^k-1$ different bitstrings 
$\mathbf{h}^i\in\set{0,1}^n\backslash\{0^n\}\, (2\le i\le 2^k)$, the subspace
$\operatorname{span}_{\mathbb{C}}\{\Ket{\mathbf{h}^1}_G,\ldots,|\mathbf{h}^{2^k}\rangle_G\}$,
where $\mathbf{h}^1=0^n$, is a $[[n,k,d]]$ QECC iff 
\begin{align}
\mathbf{h}^i\in Z^\perp(G,n,d) ~\forall i\in [2^k],\label{eq:condition1} \\
\mathbf{h}^i+\mathbf{h}^j\notin W (G,n,d) ~\forall 1\le i\ne j\le 2^k.\label{eq:condition2}
\end{align}
\end{corollary}
\noindent If $\operatorname{span}_{\mathbb{C}}\{\Ket{\mathbf{h}^1}_G,\ldots,
|\mathbf{h}^{2^k}\rangle_G\}$ is a $[[n,k,d]]$ QECC, then $\mathbf{h}^i\in C(G,n,d)$ for every 
$i\ge 2$.
However, given a non-empty set $C(G,n,d)$ for a given graph $G$, one cannot 
claim that the subspace
\begin{equation}
\operatorname{span}_{\mathbb{C}}\left\{\Ket{G},\Ket{\mathbf{h}^1}_G,\Ket{\mathbf{h}^2}_G,\cdots \right\},    
\end{equation}
where $\mathbf{h}^i\in C(G,n,d)$, is a QECC with distance $d$; it is possible that 
$\mathbf{h}^i+\mathbf{h}^j\in W(G,n,d)$ which would violate the second QECC condition, 
Eq.~(\ref{eq:condition2}). To construct a $[[n,k,d]]$ QECC with $k>1$, one 
needs to carefully choose a subset of $C(G,n,d)$ to ensure that both conditions,
Eqs.~(\ref{eq:condition1}) and~(\ref{eq:condition2}) are simultaneously 
satisfied.

For an arbitrary given graph and a set of bitstrings $\set{\mathbf{h}^i}$, it may not be 
straightforward to verify the conditions in 
Corollary~\ref{corollary:QECCgeneralconditionZWset}. But by making use of the 
special structure of the given graph, it is often possible to obtain some 
useful results, as discussed in detail in the next section.

\section{Examples}
\label{sec:ExampleAnalysis}
In this section, we consider various examples of graph-state families. Using 
the conditions derived in the last section, particularly 
Theorem~\ref{theorem:TQO1graphstate}, we show which families are in TQO-1 and 
which are not, leading to both topologically trivial and non-trivial states,
respectively. Examples of topologically trivial graph states include regular 
lattices in arbitrary dimensions, as well as the star and complete graphs 
(which are equivalent to one another); graph states in TQO-1 include the state 
associated with the toric code graph, the line graph of the complete graph, the 
line graph of the complete bipartite graph, and a generalized toric graph. 
These particular graphs are given as examples because their connectivity is 
simply described for arbitrary sizes, and they provide contrasting cases for 
the existence of a graph family to exhibit topological order.

Before considering specific examples, it is important to mention that one does 
not need to know the exact value of $d^{\rm max}$ to decide whether a given 
family of graph states is in TQO-1 or is topologically trivial, despite the 
statement 
in Theorem~\ref{theorem:TQO1graphstate}. For example, if one can prove that
$C(G,n,d)=\emptyset$ for some $d\in O(1)$, then immediately one also knows that 
$d^{\rm max}\in O(1)$ because $d^{\rm max}<d$; thus, the corresponding family of
graph states is not in TQO-1. In contrast, if $C(G,n,d)\ne\emptyset$ for some 
$d\in \operatorname{sublin}(n)$, then $d^{\rm max}\in\operatorname{sublin}(n)$ 
because $d^{\rm max}\ge d$; thus, the corresponding family of graph states is in
TQO-1.

If it is not possible to infer the value of $d^{\rm max}$ directly from the 
graph properties, then one can nevertheless use the bisection method to 
identify $d^{\rm max}$ by brute force. First set $d=\lfloor n/2 \rfloor$ and 
decide if $C(G,n,d)=\emptyset$ by enumerating and checking every possible 
bitstring (note that this task scales exponentially in~$n$). If 
$C(G,n,\lfloor n/2 \rfloor)=\emptyset$, then $d^{\rm max}<\lfloor n/2\rfloor$; 
whereas if $C(G,n,\lfloor n/2\rfloor)\ne\emptyset$, then 
$d^{\rm max}\ge\lfloor n/2\rfloor$. Repeating the procedure about $\log(n)$ 
times, one obtains the value of $d^{\rm max}$. 

In practice, knowing specific details about the graph can significantly 
streamline the process of determining if $C(G,n,d)=\emptyset$. The calculation 
of the set $Z^\perp(G,n,d)$ does not require knowledge of all the bitstrings in 
$Z(G,n,d)$. Rather, a maximum independent subset of $Z(G,n,d)$ is sufficient to 
calculate $Z^\perp(G,n,d)$: every bitstring in $Z(G,n,d)$ are written as a 
linear combination of bitstrings from such a subset. For example, a maximum 
independent subset of $\{0^n,\mathbf{k},\mathbf{l},\mathbf{k}+\mathbf{l}\}$ is $\{\mathbf{k},\mathbf{l}\}$. In fact, it may be 
sufficient to identify a sufficiently large independent subset of $Z(G,n,d)$ 
rather than a maximum one. If $Z_s\subseteq Z(G,n,d)$ is an independent subset,
then $Z^\perp(G,n,d)\subseteq Z_s^\perp$; if 
$Z_s^\perp\backslash W(G,n,d)=\emptyset$, then $C(G,n,d)$ is also empty. This 
technique will be used often in the analyses of various graphs.

\subsection{Topologically trivial graph states}
\label{sec:trivialgraphstate}
\subsubsection{Graphs with constant degree}
In this section we show that all states represented by constant-degree graphs
are not in TQO-1; rather, $d^{\rm max}$ is upper bounded by the graph degree 
$\Delta(G)$. Notice that, for every basis bitstring $\mathbf{b}^i$:
\begin{equation}
\text{wt}(A\cdot \mathbf{b}^i\vee \mathbf{b}^i) = \Delta(G)+1.
\end{equation}
Then $\mathbf{b}^i\in Z(G,n,\Delta(G)+2)$ for every basis bitstring, where 
$d=\Delta(G)+2$. Evidently the $\mathbf{b}^i$ span all possible bitstrings, so
$Z^\perp(\Delta(G)+2)=\{0^n\}$ and likewise $C(\Delta(G)+2)=\emptyset$, so 
$d^{\rm max}\le \Delta(G)+1$. If a family of states is represented by graphs 
with constant degree, then $d^{\rm max}\in O(1)$ and it is not in TQO-1.

\begin{figure}[t]
    \centering
    \includegraphics[width=0.95\columnwidth]{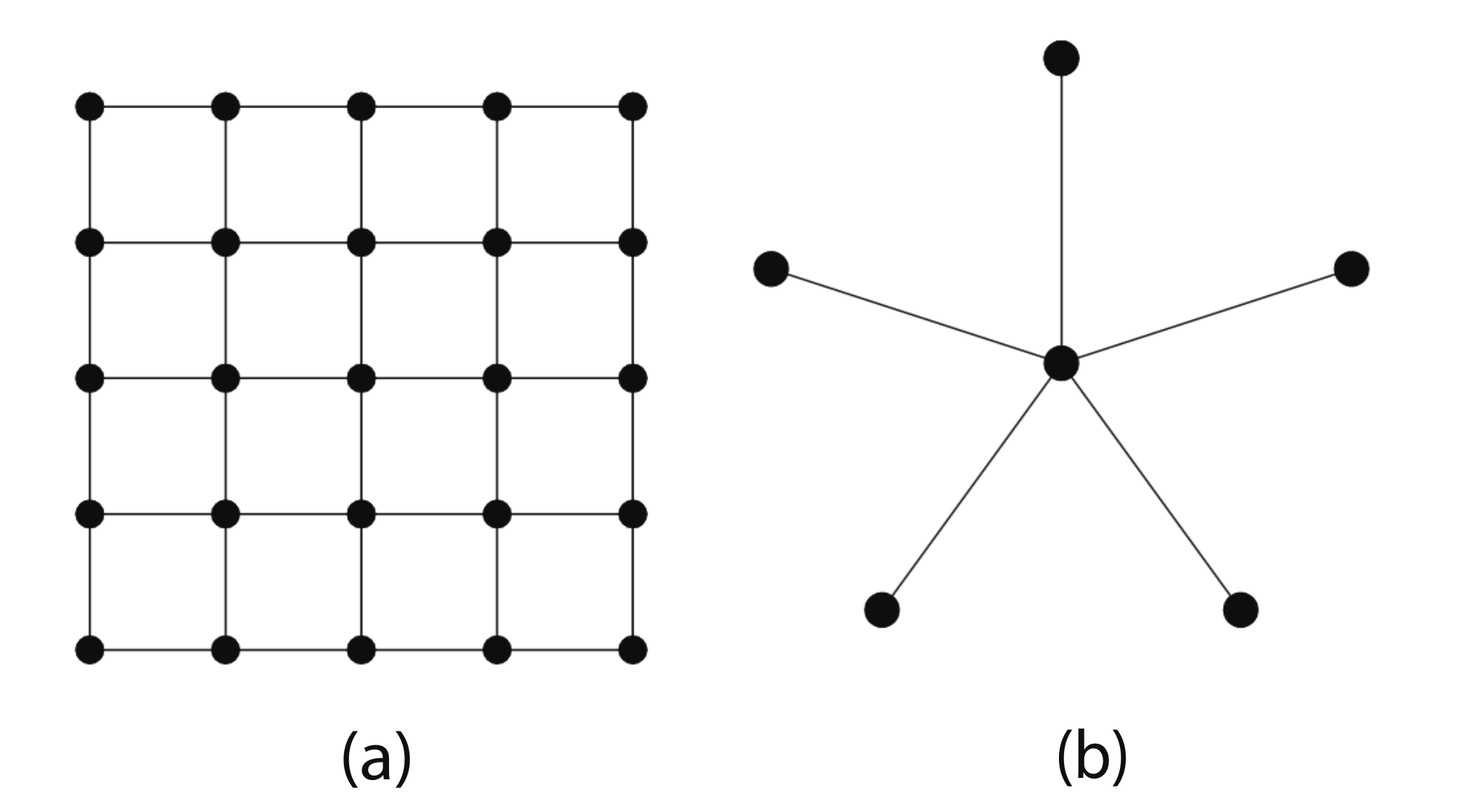}
    \caption{Examples of graphs that are not in TQO-1: (a) 2D regular lattice
    and (b) six-vertex star graph.}
    \label{fig:ExampleNonTQO1Graph}
\end{figure}

For example, consider the $D$-dimensional square lattice of linear size $L$, 
which has $L^D$ vertices, as shown in Fig.~\ref{fig:ExampleNonTQO1Graph}(a). The 
associated cluster states have symmetry-protected topological 
order~\cite{Son2011}, but they are not in TQO-1. The vertex degree 
is $\Delta(G)=2D$ regardless of the size of the lattice, and the maximum code 
distance is at most $2D+1$, which is a constant for fixed dimension even if $L$ 
increases.

From above discussion, a necessary condition for a family of graph states to be 
in TQO-1 is that the graph degree has to be macroscopic:
$\Delta(G)\in\operatorname{sublin}(n)$. This requirement is not sufficient, 
however, because there are families of macroscopic-degree graphs that are not 
in TQO-1. 
Examples include the star graph and complete graph discussed in the next two 
sections.

\subsubsection{Star graph}
\label{sec:star_graph}
Consider next the star graph on~$n$ vertices $K_{1,n-1}$, as shown in 
Fig.~\ref{fig:ExampleNonTQO1Graph}(b), which defines GHZ states on~$n$ qubits. 
The adjacency matrix, where the first vertex is chosen to have high degree, is
\begin{equation}
\label{eq:adjacencystar}
A_{1,n-1} = 
\begin{bmatrix}
0 & 1 & 1 & \cdots 1\\
1 & 0 & 0 & \cdots 0\\
\vdots & \vdots & \vdots & \cdots \vdots\\
1 & 0 & 0 & \cdots 0
\end{bmatrix}.
\end{equation}
When $i>1$, $\text{wt}\left[\left(A_{1,n-1}\cdot \mathbf{b}^1\right)\vee \mathbf{b}^1\right]=n$ and
$\text{wt}\left[\left(A_{1,n-1}\cdot \mathbf{b}^i\right)\vee \mathbf{b}^i\right]=2$. The second of these
yields $\left\{\mathbf{b}^i\,(i>1)\right\}\subseteq Z(G_{\rm star},n,3)$ and 
$$Z(K_{1,n-1},n,3)^\perp\subseteq\left\{\mathbf{b}^i\,(i>1)\right\}^\perp=\{0^n,\mathbf{b}^1\}.$$
However, the bitstrings $\{0^n,\mathbf{b}^1\}\subseteq W(G_{\rm{star}},n,3)$ because 
their Hamming weights are smaller than 3. Thus, 
$C(K_{1,n-1},n,3)=\emptyset$ and $d^{\rm max}<3$, and the family of star graph 
states is therefore not in TQO-1.  

\subsubsection{Complete graph}
\label{sec:complete_graph}
Consider next the complete graph on~$n$ vertices, $K_n$, which is 
local-Clifford equivalent (LC-equivalent) to the star 
graph~\cite{Van2004Graphical}. An example is shown in 
Fig.~\ref{fig:completegraph}(a). 
The adjacency matrix is $A_K=J-I$, where $J$ is the matrix of all ones. In this 
case, $\text{wt}\left[\left(A_K\cdot \mathbf{b}^i\right)\vee \mathbf{b}^i\right]=n\,\forall i$, which 
would imply that $d=n+1$, an impossibility. Next consider weight-two 
bitstrings, corresponding to the $n-1$ independent pairings of the basis 
states, $\mathbf{c}^{i}=\mathbf{b}^i+\mathbf{b}^{i+1}$, $i<n$. Then $\text{wt}\left(A_K\cdot \mathbf{c}^i\right)
=w\left[\left(A_K\cdot \mathbf{c}^i\right)\vee \mathbf{c}^i\right]=2$, so that
$\left\{\mathbf{c}^1,\ldots,\mathbf{c}^{n-1}\right\} \subseteq Z(K_n,n,3)$. Only the all-zero
and all-one bitstrings are orthogonal to $\mathbf{c}^i\; (i<n)$, so that 
$$Z^\perp(K_n,n,3)\subseteq \left\{\mathbf{c}^1,\ldots,\mathbf{c}^{n-1}\right\}^\perp 
= \{0^n,1^n\}. $$
But as was the case for star graphs, both $0^n$ and $1^n$ are also in 
$W(K_n,n,3)$ because $1^n=A_K\cdot \mathbf{b}^i+\mathbf{b}^i\,\forall i$.
One therefore again obtains $C(K_n,n,3)=Z^\perp(K_n,n,3)\backslash W(K_n,n,3)
=\emptyset$, and $d^{\rm max}<3$ for the family of complete graph states. The 
results for the star and complete graphs are consistent with the fact that the 
GHZ state is not topologically ordered.

\subsection{Topologically non-trivial graph states and codes}
\label{sec:TopoGraphCode}
In this section, we present some topologically non-trivial graph states and 
construct their associated QECCs with macroscopic distance. By embedding a 
classical LDPC code into the multi-star graph state, we obtain an $[[n,k,d]]$ 
QECC where $n=\Theta(kd)$. Two $[[n,1,\Theta(\sqrt{n})]]$ QECCs are based on the 
line graphs of the complete and completely bipartite graphs, corresponding to
the triangular and rook's graph, respectively. 

Despite the macroscopic distance of the QECCs presented here, it might be 
considered an abuse of notation to refer to them as topological codes. They do 
not necessarily all have local stabilizer generators, and it is not clear if
the degeneracy of the code subspace has a topological origin like is the case 
for the 2D toric code and color code. Nevertheless, the constructions presented
below are all rather simple while possessing a code distance that scales 
sublinearly in the number of physical qubits. Thus, we hope that these examples
might be helpful in the understanding of how to construct general topological 
states and QECCs.

\begin{figure}[t]
    \centering
    \includegraphics[width=0.95\columnwidth]{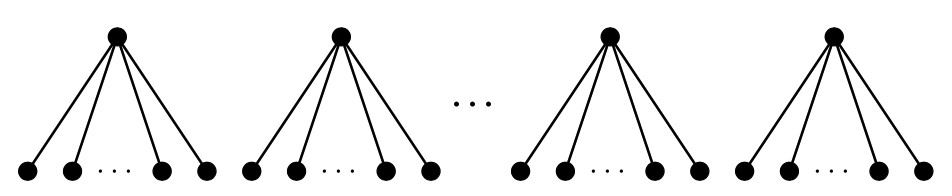}
    \caption{Multiple copies of the star graph}
    \label{fig:MulStar}
\end{figure}

\subsubsection{Multiple copies of the star graph}
\label{sec:multistargraph}
Perhaps surprisingly, multiple copies of star graphs lead to a different 
conclusion than was found for a single copy of the star graph. Consider an
$n=qm$-vertex graph $G_{\rm{mstar}}$ composed of $q$ disconnected components, 
each of which is a star graph on $m$ vertices and $q\ge m$. The adjacency 
matrix of this graph is block diagonal, $A_{\rm mstar}=A_{1,m-1}^{\oplus q}$, 
where $A_{1,m-1}\in\mathbb{Z}_2^{m\times m}$ is of same form as in 
Eq.~(\ref{eq:adjacencystar}). For this family of graph states, 
$d^{\rm max}=m=n/q$, which scales macroscopically with~$n$.

As in the case of a single star graph, $\text{wt}(A_{\rm mstar}\cdot \mathbf{b}^i)=1$ and 
$\text{wt}\left[(A_{\rm mstar}\cdot \mathbf{b}^i)\vee \mathbf{b}^i\right]=2$ when 
$(i\,{\rm mod}\,m)\neq 1$, so that 
$\left\{\mathbf{b}^i| (i\,{\rm mod}\,m)\ne 1\right\}\subseteq Z(G_{\rm{mstar}},n,3)$.
Furthermore, $\left\{\mathbf{b}^i| i\mod m\ne 1\right\}$ is the maximum independent 
subset of $Z(G_{\rm{mstar}},n,m)$, because any linear combination of these 
bitstrings will have maximum weight $m-1$ but each only has the first vertex as
a neighbor. For any bitstring $\mathbf{k}$ with a 1 at position $i\mod m=1$, 
$\text{wt}[(A\cdot \mathbf{k})\vee \mathbf{k}] \ge m$ regardless the bit value in other positions. 
Furthermore, all linear combinations of $\mathbf{b}^i$ with $(i\,{\rm mod}\,m)=1$ are 
orthogonal to bitstrings in $Z(G_{\rm{mstar}},n,m)$:
\begin{equation}
Z^\perp(G_{\rm{mstar}},n,m) = \operatorname{span}_{\mathbb{Z}_2}
\left\{\mathbf{b}^i|(i\,{\rm mod}\,m)=1\right\}.
\end{equation}

For TQO-1, $Z^\perp(m)\backslash W(m)$ must not be an empty set; in the present
case, we show that every bitstring in $Z^\perp(m)$ with Hamming weight greater 
or equal to $m$ is not in $W(m)$. 

Consider first the conditions where $h=A\cdot \mathbf{k}+\mathbf{l}=\mathbf{b}^1$: $(A\cdot \mathbf{k})_1=0$ and 
$\mathbf{l}_1=1$, and $(A\cdot \mathbf{k})_1=1$ and $\mathbf{l}_1=0$. In both cases, $\text{wt}(\mathbf{k}\vee \mathbf{l})\ge 
\text{wt}(A\cdot \mathbf{k}+\mathbf{l})=1$. Next, consider the case where $h=\mathbf{b}^1+\mathbf{b}^{m+1}$. 
Similarly, either $(A\cdot \mathbf{k})_{m+1}=0$ and $\mathbf{l}_{m+1}=1$, or $(A\cdot \mathbf{k})_{m+1}=1$
and $\mathbf{l}_{m+1}=0$. Thus, $\{(A\cdot \mathbf{k})_1,\mathbf{l}_1,(A\cdot \mathbf{k})_{m+1},\mathbf{l}_{m+1}\}
=\{1,0,1,0\},\;\{1,0,0,1\},\; \{0,1,1,0\}$, and $\{0,1,0,1\}$.
In the first case, $(A\cdot \mathbf{k})_1=(A\cdot \mathbf{k})_{m+1}=1$, so there are an odd 
number of indices $i\in [2,m]$ where $\mathbf{k}_i=1$ and an odd number of indices
$j\in [m+2,2m]$ where $\mathbf{k}_j=1$; therefore $\text{wt}(\mathbf{k}\vee \mathbf{l})\ge 2$.
In the second case, there are again an odd number of indices $i\in [2,m]$ where 
$\mathbf{k}_i=1$; together with $\mathbf{l}_{m+1}=1$, one again obtains $\text{wt}(\mathbf{k}\vee \mathbf{l})\ge 2$. 
Likewise for the third case. Finally, $\text{wt}(\mathbf{k} \vee \mathbf{l})\ge \text{wt}(\mathbf{l})\ge 2$ in the fourth
case because of $\mathbf{l}_1=\mathbf{l}_{m+1}=1$. Every non-zero entry in $A\cdot \mathbf{k}+\mathbf{l}$ either 
requires an entry in $l$ or $\mathbf{k}$ to be non-zero. Generalizing the above argument 
to arbitrary linear combinations of bitstrings from 
$\left\{\mathbf{b}^i|(i\,{\rm mod}\,m)=1\right\}$, one obtains
$\text{wt}(\mathbf{k}\vee \mathbf{l})\ge \text{wt}(\mathbf{h})$ if $\mathbf{h}\in Z^\perp$. Therefore, if the
$\text{wt}(\mathbf{h})\geq m$ then $\text{wt}(\mathbf{k}\vee \mathbf{l})\ge m$, i.e.\ $\mathbf{h}\notin W(m)$; thus
$C(G_{\rm mstar},n,m)\ne\emptyset$.

Increasing $d$ to $m+1$ would result in $\mathbf{b}^i\in Z(G_{\rm{mstar}},n,m+1)$ when 
$i=1\mod m$, because $\text{wt}(A\cdot \mathbf{b}^i\vee \mathbf{b}^i )=m\le m+1$. In that case, every 
basis bitstring is in $Z(m+1)$ so that $Z^\perp(m+1)=\{0^n\}$ and 
$C(m+1)=\emptyset$. One therefore concludes that for the family of multi-star 
graph states, $d^{\rm max}=m$. If $q=\Theta(m)$, then 
$d^{\rm max}=\Theta(\sqrt{n})$ and this family of graph states is in TQO-1.
The same conclusion evidently also holds for multiple copies of complete 
graphs.

In summary, for an $n=qm$-vertex multi-star graph, composed of $q$ disconnected 
$m$-vertex star graphs ($q\ge m$):
\begin{eqnarray}
C(G_{\rm mstar},n,m)&=&\{\mathbf{h}\in
\operatorname{span}_{\mathbb{Z}_2}\left\{\mathbf{b}^i|(i\,{\rm mod}\,m)=1\right\}
\nonumber \\
&&\quad | \text{wt}(\mathbf{h})\ge m\}.
\end{eqnarray}
Combining this result with Corollary~\ref{corollary:QECCgeneralconditionZWset}, 
one can construct a $[[qm,\Theta(q),m]]$ QECC using a classical LDPC code.
Suppose one is given a classical linear code $\mathtt{C}=[q,c_1q,c_2q]$, where 
$c_1$ and $c_2$ are constant and $c_2q\ge m$. Each codeword in this classical 
code are mapped to a logical state in a QECC in the following way:
\begin{eqnarray}
\mathbf{h}\in \mathtt{C}\subseteq\{0,1\}^q &\mapsto& 
r(\mathbf{h})=\sum_{i\in[q]} \mathbf{h}_i \mathbf{b}^{1+m(i-1)}\in\set{0,1}^n\nonumber\\
&\mapsto& \Ket{r(\mathbf{h})}_{G_{\rm mstar}} \in\mathscr{H}_2^{\otimes n}.
\end{eqnarray}
As the Hamming weight of every bitstring in $\mathtt{C}$ is no less than 
$c_2q$, every $r(\mathbf{h})$ mapped from the classical code also has Hamming weight no 
less than $c_2q$.
\begin{theorem}
Subspace $\operatorname{span}_{\mathbb{C}}\{ \Ket{r(\mathbf{h})}_{G_{\rm mstar}}\}$ is a $[[qm,c_1q,m]]$ QECC.
\end{theorem}
\begin{proof}
From Corollary~\ref{corollary:QECCgeneralconditionZWset}, one only need show 
that the following conditions hold for all $\mathbf{h}\ne \mathbf{h}^\prime \in \mathtt{C}$:
\begin{align}
\label{eq:r(h)}
r(\mathbf{h}) \in    C(G_{\rm mstar},n,m);\\
\label{eq: r(h)r(h')}
r(\mathbf{h}) + {r(\mathbf{h}^\prime) } \notin W(G_{\rm mstar},n,m).
\end{align}
Following the arguments presented in Sec.~\ref{sec:multistargraph}, 
Eq.~\ref{eq:r(h)} holds because $r(\mathbf{h})$ is a linear combination of 
$\left\{\mathbf{b}^i|(i\,{\rm mod}\,m)=1\right\}$. In the classical LDPC code 
$\mathtt{C}$, one has $\text{wt}(h+\mathbf{h}^\prime)\ge c_2q$ for every 
$h,\mathbf{h}^\prime\in\mathtt{C}$. Therefore,
\begin{equation}
\text{wt}(r(\mathbf{h})+r(\mathbf{h}^\prime))\ge c_2q, \forall \mathbf{h},\mathbf{h}^\prime\in\mathtt{C},
\end{equation}
leading to Eq.~(\ref{eq: r(h)r(h')}).
\end{proof}

\subsubsection{Toric graph}
\label{sec:toricgraph}
\begin{figure}[t]
    \centering
    \includegraphics[width=0.95\columnwidth]{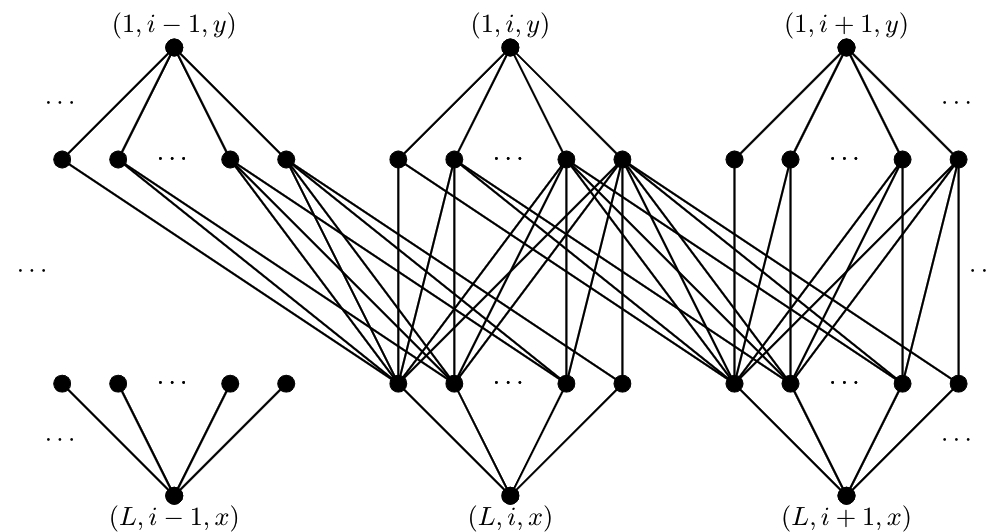}
    \caption{Toric graph, { where $(i,j,x/y)$ is the coordinate of the qubits in the 2D toric code as in Sec.~\ref{sec:toricgraph}}}
    \label{fig:ToricGraph}
\end{figure}

The toric graph is a $2L^2$-vertex graph representing a graph state, and is
LC-equivalent to one of the ground states in the toric 
code~\cite{liao2021graph}. The toric graph state is in TQO-1 because the toric 
code has macroscopic distance $d=L\in O(\sqrt{n})$. Nevertheless, it might be
helpful to the reader to confirm this result using the method discussed in
this work, by showing that set $C(G_{\rm toric},2L^2,L)$ is non-empty. The 
qubits in the 2D toric code are placed on the edges of a square 
$L\times L$ lattice. Qubits are labelled by 
$(i,j,d)\in[L]\times[L]\times\{x,y\}$, where $(i,j)$ are the spatial 
coordinates and $x$ and $y$ denote the orientation of the qubit on horizontal 
or vertical edges. The adjacency matrix of the toric graph 
is~\cite{liao2021graph}
\begin{eqnarray}
A_{ijd_1,lmd_2}&=&\delta_{d_1,x}\delta_{d_2,x}\delta_{m,j}
(\delta_{l,L}\theta_{i,L-1}+\delta_{i,L}\theta_{l,L-1})\nonumber \\
&+&\delta_{d_1,y}\delta_{d_2,y}\delta_{m,j}
(\delta_{i,1}\theta_{2,l}+\delta_{l,1}\theta_{2,i})\nonumber \\
&+&\delta_{d_1,y}\delta_{d_2,x}(\delta_{m,j}
+\delta_{m-1,j})\theta_{l,i-1}\theta_{2,i}\nonumber \\
&+&\delta_{d_1,x}\delta_{d_2,y}(\delta_{m,j}
+\delta_{j-1,m})\theta_{i+1,l}\theta_{i,L-1},\hphantom{aa}
\label{eq:toric_adjacency_matrix}
\end{eqnarray}
where $\delta_{i,j}$ and $\theta_{i,j}$ are the usual Kronecker and Heaviside
theta functions, respectively:
\begin{equation}
\delta_{i,j}=\begin{cases}
1 & j=i\\
0 & {\rm otherwise}
\end{cases}\;{\rm and}\;
\theta_{i,j}=\begin{cases}
1 & i\leq j\\
0 & {\rm otherwise}.
\end{cases}
\end{equation}
While it might not be immediately apparent from the form of the adjacency 
matrix, the toric graph in fact corresponds to multiple copies of the star 
graph, which are then connected by half graphs~\cite{liao2021graph}, as shown in 
Fig.~\ref{fig:ToricGraph}. Appendix~\ref{appendix:ToricGraph} shows that
\begin{equation}
C(G_{\rm toric},2L^2,L) = \left\{\sum_{j=1}^{L} \mathbf{b}^{Ljx},\sum_{i=1}^{L} \mathbf{b}^{1jy},\sum_{i=1}^{L} ( \mathbf{b}^{Ljx}+ \mathbf{b}^{1jy})\right\}.     
\end{equation}
These three bitstrings correspond to the three graph basis states that are 
LC-equivalent to 
the logical states of the toric code. Together with the toric graph state 
itself (the zero state in this representation), they span a subspace that is 
locally equivalent to the 2D toric code.

\subsubsection{Connected multiple star graphs}
\label{sec:connmulstar}
\begin{figure}[t]
    \centering
    \includegraphics[width=0.95\columnwidth]{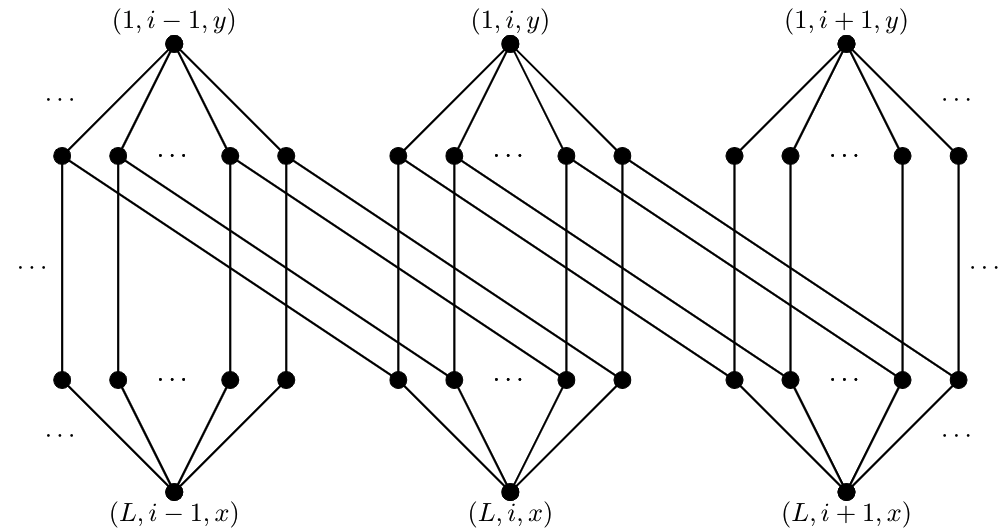}
    \caption{Multi-star graph connected by constant-degree subgraphs, { where $(i,j,x/y)$ is the coordinate of the qubits as in Sec.~\ref{sec:connmulstar}}}
    \label{fig:ConnectedMulStar}
\end{figure}

Given that both multiple star graphs and the toric graph (multiple star graphs 
connected via half graphs) are in TQO-1, one might wonder if any connected graph 
$G_{\rm cmstar}$ that is formed by a regular linking of multiple star graphs is
also in TQO-1. It turns out that this is in fact the case, as is proven here. 
An example $G_{\rm cmstar}$ is depicted in Fig.~\ref{fig:ConnectedMulStar}.

As was the case for the toric graph, label the vertices as 
$(i,j,d)\in[L]\times[L]\times\{x,y\}$. There are $2L(L-1)$ vertices in the 
middle two layers that are labelled by $[2,L]\times [L]\times \{y\}$ and  
$[L-1]\times [L]\times \{x\}$. As the corresponding vertices have constant 
degree $\text{wt}(A\cdot \mathbf{b}^{ijd})$ (which is 3 in the example shown in
Fig.~\ref{fig:ConnectedMulStar}), one obtains
$\mathbf{b}^{ijd}\in Z(G_{\rm cmstar},2L^2,L)$ if 
$(i,j,d)\in [2,L]\times [L]\times \{y\} \cup [L-1]\times [L]\times \{x\}$.
On the other hand, the neighborhood of vertices $(1,i,y)$ and $(L,j,x)$, 
$i,j\in[L]$, is composed of $L-1$ vertices, and each of these neighborhoods is
disjoint. As a result, if bitstring $\mathbf{k}$ has non-zero entry at $(1,i,y)$ or 
$(L,j,x)$, then $\text{wt}(\mathbf{k}\vee A \mathbf{k})\ge L$ and so 
$\mathbf{b}^{1iy},\mathbf{b}^{Ljx} \notin Z(G_{\rm cmstar},2L^2,L)$ for every $i,j\in[L]$. Thus
$Z^{\perp}(L) =\operatorname{span}\{\mathbf{b}^{11y},\cdots,\mathbf{b}^{1Ly},\mathbf{b}^{L1x},\cdots,
\mathbf{b}^{LLx}\}$ and 
\begin{equation}
\left\{\sum_{j=1}^{L} \mathbf{b}^{Ljx},\sum_{i=1}^{L} \mathbf{b}^{1jy},\sum_{i=1}^{L} ( \mathbf{b}^{Ljx}+ \mathbf{b}^{1jy})\right\} \subseteq Z^\perp(L).
\end{equation}
These three bitstrings are not in $W(L)$ for reasons similar to those discussed 
in Sec.~(\ref{sec:multistargraph}), so one obtains the result that 
$C(G_{\rm cmtar},2L^2,L)\neq\emptyset$ and the family of connected multi-star 
states is also in TQO-1.

\subsubsection{Line graph of the complete graph}
\label{sec:linecompletegraph}
\begin{figure}[t]
     \centering
     \includegraphics[width=0.95\columnwidth]{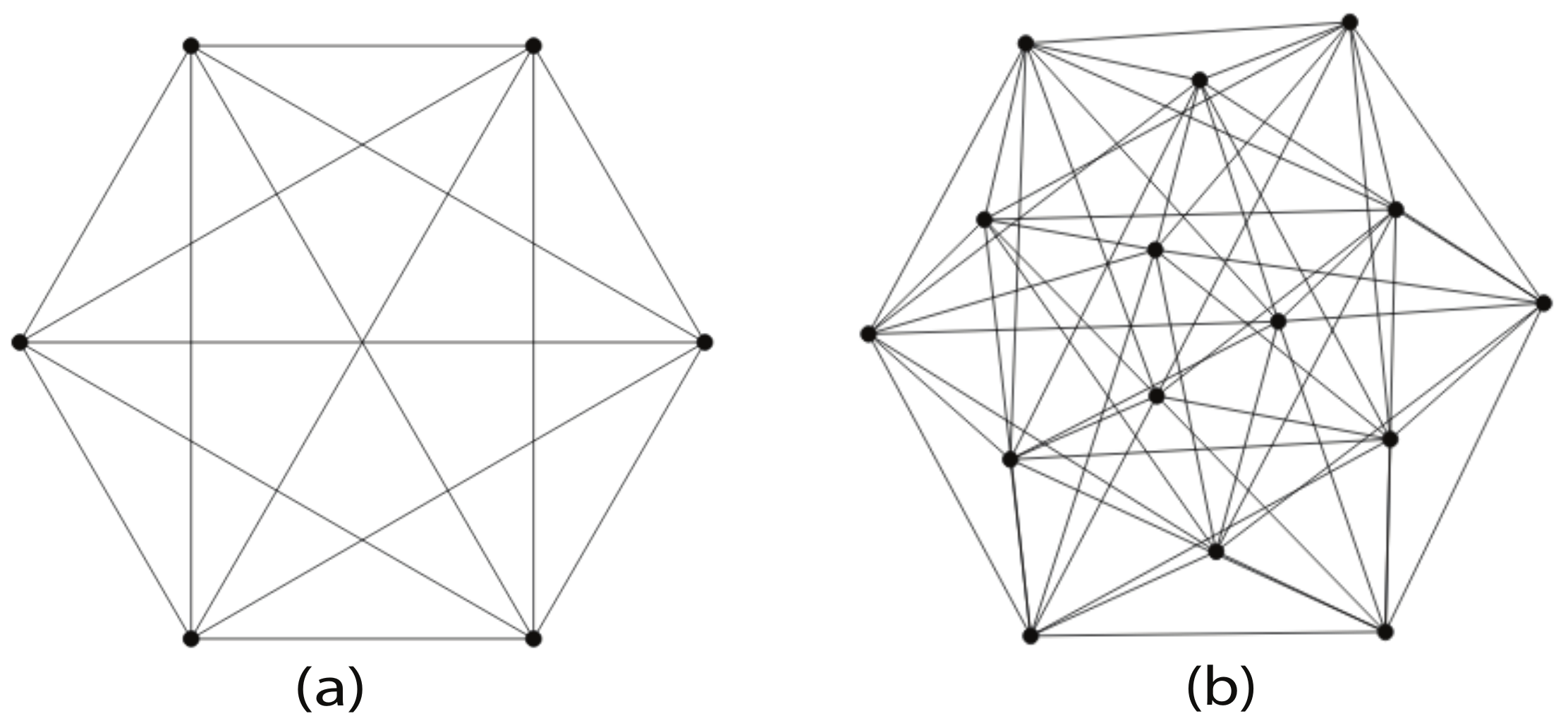}
        \caption{(a) The complete graph $K_6$; (b) the line graph of $K_6$.}
        \label{fig:completegraph}
\end{figure}

One of the distinguishing features of the previous two examples, multiple
copies of the star graph and the toric graph, is that the maximum vertex degree
(the maximum value of the vertex degree taken over all vertices) increases with 
the total number of vertices. Evidently the examples of a single copy of the 
star and complete graph show that this condition is not sufficient for a family 
of graphs to be in TQO-1. It is nevertheless worthwhile to consider other 
examples of graphs with this behaviour. One candidate is the line graph of the 
complete graph, $L(K_m)$, also known as the triangular graph $T_m$, considered 
here. An example of a complete graph and its triangular graph are depicted in 
Fig.~\ref{fig:completegraph}.

The graph $K_m=(V,E)$ is the $m$-vertex complete graph where each pair of 
vertices shares an edge, so the number of edges is 
$|E|=\binom{m}2={m(m-1)}/{2}$. The line graph of $K_m$ is 
$L(K_m)=T_m=(V^\prime,E^\prime)$; by definition, 
$|V^\prime|=|E|=n$; the number of edges is $|E^{\prime}|=2(m-2)$. The adjacency 
matrices of $K_m$ and $T_m$ are denoted $A\in\mathbb{Z}_2^{m\times m}$ and 
$A^\prime\in\mathbb{Z}_2^{n\times n}$, respectively. The main result in this 
section is Theorem~\ref{theorem:line_complete_TQO1}, which is proven by showing 
that the set $C(T_m,\binom{m}2,\lfloor m/2\rfloor)$ is non-empty for all 
$2\le m\in\mathbb{N}$.
\begin{theorem}
\label{theorem:line_complete_TQO1}
The family of line graph states $\mathtt{S}=\Set{\Ket{T_m}}_{m\ge2}$ is in 
TQO-1.
\end{theorem}
\begin{proof}
A full proof is given in Appendix~\ref{appendix:linegraphcompletegraph}, but
the main ideas are sketched here. Consider a bitstring $\mathbf{s}^v\in\set{0,1}^{|E|}$, 
which only has non-zero entries at edges incident to vertex $v$ in graph $K_m$.
From the construction of the line graph, the column vector of the adjacency 
matrix $A^\prime\cdot \mathbf{b}^{e_{ij}}$ only has non-zero entries at edges incident 
to either $v_i$ or $v_j$ (except for $e_{ij}$ itself), which are formally 
expressed as $A^\prime\cdot \mathbf{b}^{e_{ij}} =\mathbf{s}^{v_i}+\mathbf{s}^{v_j}$; then
\begin{equation}
A^\prime \cdot \mathbf{k} = \sum_{e_{ij}} \mathbf{k}_{e_{ij}}(A^\prime\cdot \mathbf{b}^{e_{ij}})  
= \sum_{e_{ij}} \mathbf{k}_{e_{ij}}(\mathbf{s}^{v_i}+\mathbf{s}^{v_j}).
\end{equation}
The sum (bitwise XOR) of $\mathbf{s}^v$ over edges are equivalently expressed as a 
sum over a subset of vertices
\begin{equation}
A^\prime \cdot \mathbf{k}  = \sum_{v\in V_\mathbf{k}^{\rm o}} \mathbf{s}^v,    
\end{equation}
where $V_\mathbf{k}^{\rm o}$ is the subset of vertices with odd degree in subgraph 
$(K_m)_\mathbf{k}$. Using this relation, one can obtain the Hamming weight 
$\text{wt}(A^\prime \cdot \mathbf{k})=l_\mathbf{k}(m-l_\mathbf{k})$, where $l_\mathbf{k}=|V_\mathbf{k}^{\rm{o}}|$. Next, one can 
show that $\text{wt}(A^\prime \cdot \mathbf{k})< \lfloor m/2\rfloor$ only if $l_\mathbf{k}=0$, i.e.\
that all the vertices in subgraph $(K_m)_{\mathbf{k}}$ have even degree. As a result, 
every bitstring $\mathbf{k}$ in $Z(T_m,\binom{m}2,\lfloor m/2\rfloor)$ corresponds to a 
subgraph 
$(K_m)_\mathbf{k}$ that is an Eulerian cycle or sum (edge symmetric difference) of 
Eulerian cycles. Because of the special structure of the complete graph, every 
cycle in $K_m$ can be decomposed into a sum of triangles, which allows for the 
determination of a maximum independent subset of $Z(\lfloor m/2\rfloor)$ with
every member a bitstring corresponding to a triangle; furthermore, the
bitstring $\mathbf{s}^v$ can be shown to be orthogonal to all of these bitstrings, so
that $\mathbf{s}^v\in Z^\perp(T_m,\binom{m}2,\lfloor m/2\rfloor)$ for every $v\in V$.
Finally, using standard algebraic arguments, one can prove that 
$\mathbf{s}^v\notin W^\perp(T_m,\binom{m}2,\lfloor m/2\rfloor)$, so that 
$\mathbf{s}^v\in C(T_m,\binom{m}2,\lfloor m/2\rfloor)$ for every $v\in V$. Given that 
$ C(T_m,\binom{m}2,\lfloor m/2\rfloor)\neq\emptyset$ and the code distance is 
macroscopic, the family of line graphs of the complete graph is in TQO-1.
\end{proof}

Because of Lemma~\ref{lemma:CGnd_qecc} and $\mathbf{s}^{v_1}\in C(L(K_m),\binom{m}2, 
\lfloor m/2\rfloor)$, the following result holds:
\begin{corollary}
Subspace $\operatorname{span}_{\mathbb{C}} \Set{\Ket{L(K_m)},
\Ket{\mathbf{s}^{v_1}}_{L(K_m)}}$ is a $[\binom{m}2,1, \lfloor m/2\rfloor]$ QECC.
\end{corollary}
An immediate question is: is it possible to expand this code using 
Corollary~\ref{corollary:QECCgeneralconditionZWset}, so that it can encode more 
than one logical qubit while simultaneously keeping the same code distance 
$d=\lfloor m/2\rfloor$. It turns out that this is not possible, unfortunately. 
Given that $C(L(K_m),\binom{m}2, \lfloor m/2\rfloor)$ contains more than one 
bitstring, one would be tempted to choose another one, for example $\mathbf{s}^{v_2}$, 
so that
$$\operatorname{span}_{\mathbb{C}}\Set{\Ket{L(K_m) },\Ket{\mathbf{s}^{v_1}}_{L(K_m)},
\Ket{\mathbf{s}^{v_2}}_{L(K_m)}}$$ 
is also a QECC with distance $d=\lfloor m/2\rfloor$. However, from 
Corollary~\ref{corollary:QECCgeneralconditionZWset}, for the expanded 
three-dimensional subspace to be a QECC with $d=\lfloor m/2\rfloor$, $\mathbf{s}^{v_2}$ 
needs to be in $Z^\perp( \lfloor m/2\rfloor)$ while $\mathbf{s}^{v_1}+\mathbf{s}^{v_2}$ cannot be
in $W$. The second condition does not hold because $\mathbf{s}^{v_1}+\mathbf{s}^{v_2}$ is one of 
the column vectors of the adjacency matrix $A^\prime$.
For any other bitstring $\mathbf{h}\in C(L(K_m),\binom{m}2,\lfloor m/2\rfloor)$, similar 
problems will be present, so that it is not possible to expand 
$\operatorname{span}_{\mathbb{C}}\{\Ket{L(K_m) },\Ket{\mathbf{s}^{v_1}}_{L(K_m)}\}$ to 
encode more logical qubits without sacrificing the code distance.

\subsubsection{Line graph of the complete bipartite graph}
\label{sec:LineCompleteBiGraph}
\begin{figure}[t]
     \centering
\includegraphics[width=0.95\columnwidth]{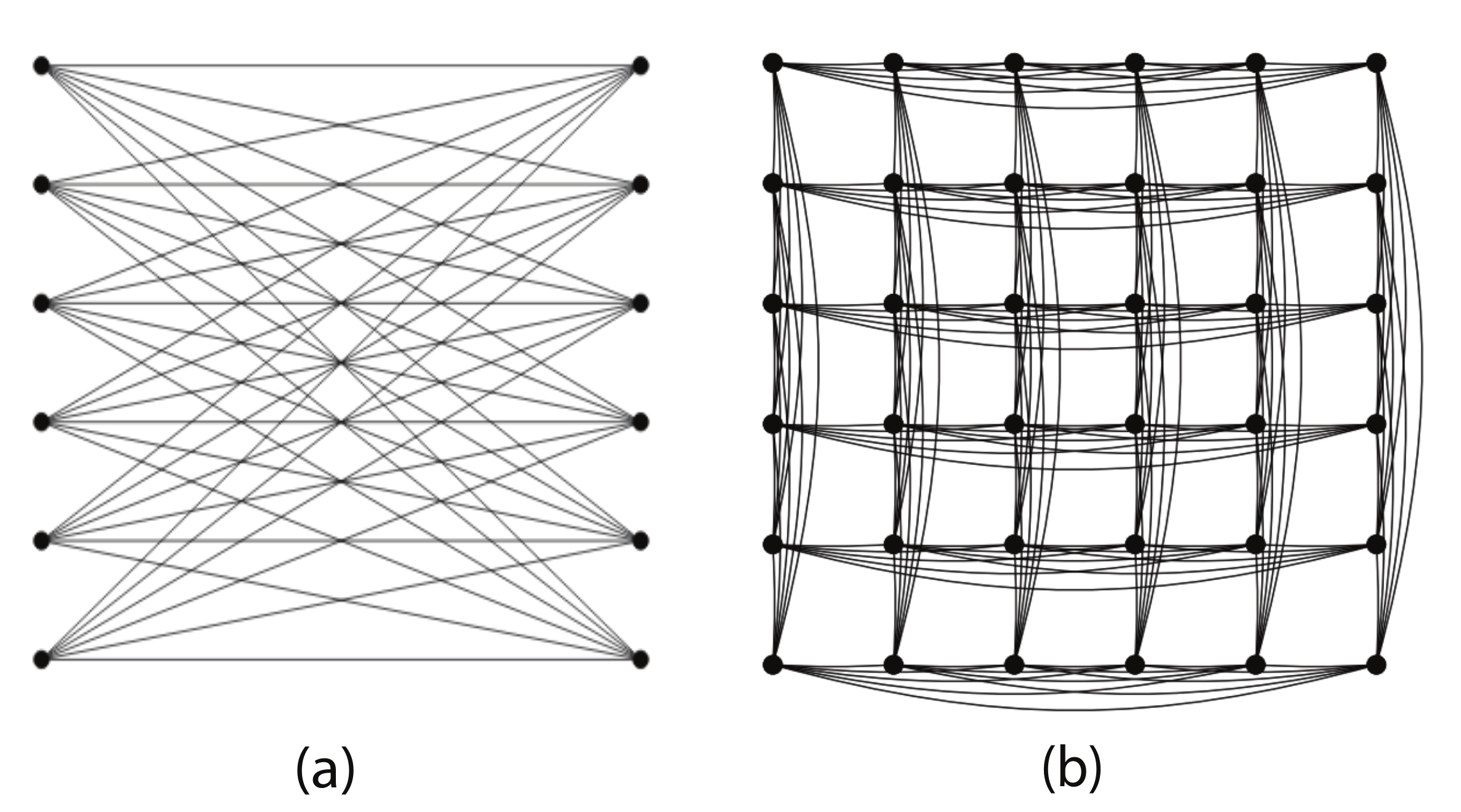}
        \caption{(a) The complete bipartite graph $K_{6,6}$ and (b) the line 
	graph of $K_{6,6}$.}
        \label{fig:completebipartitegraph}
\end{figure}

The second example considered here is similar to the first one: the graph state 
represented by the line graph of the balanced complete bipartite graph 
$L(K_{m,m})$, which is also called a rook's graph because the edges represent
all the possible moves a rook can take in the game of chess. An example of a
complete bipartite graph and its associated rook's graph are depicted in 
Fig.~\ref{fig:completebipartitegraph}.

The complete bipartite graph $K_{m,m}$ has $2m$ vertices divided into two 
complementary subsets of the same size 
$V=X\sqcup Y=\set{v^x_1,\ldots,v_m^x}\sqcup \set{v^y_1,\ldots,v_m^y}$.
Similarly, $V,E,A$ denote the set of vertices, the set of edges, and the 
adjacency matrix of $K_{m,m}$, respectively, while $V^\prime,E^\prime,A^\prime$ 
are used for the line graph $L(K_{m,m})$. In $K_{m,m}$, every vertex in the
set $X$ is connected to every vertex in the $Y$, but there are no edges
connecting two vertices in $X$ or two vertices in $Y$; there are therefore 
$m^2$ edges in total: $|E|=m^2$ and $|V^\prime|=n=m^2$.

The main result of this section is Theorem~\ref{theorem:line_complete_bi_TQO1}, 
which is proven by showing that $C(L(K_{m,m}),m^2, m)\ne \emptyset$:
\begin{theorem}
\label{theorem:line_complete_bi_TQO1}
The family of line graph states $\mathtt{S}=\Set{\Ket{L(K_{m,m})}}_{m\ge4}$ is 
in TQO-1.
\end{theorem}
\begin{proof}
Again, only a sketch of the proof is provided here, and the full proof is found
in Appendix~\ref{appendix:linecompletebigraph}. Analogously, $\mathbf{s}^v$ represents 
the bitstring that only has non-zero entries at edges incident to $v$ in graph 
$K_{m,m}$. The column vector $A^\prime\cdot \mathbf{b}^e$, where $e=(v_i^x,v_j^y)$, only 
has non-zero entries at edges incident to either $v_i^x$ or $v_j^y$, so that
$A^\prime\cdot \mathbf{b}^e = \mathbf{s}^{v_i^x}+\mathbf{s}^{v_j^y}$. Similarly, 
$A^\prime \cdot \mathbf{k} = \sum_{v\in V_\mathbf{k}^{\rm o}} \mathbf{s}^v$, where $V_\mathbf{k}^{\rm{o}}$ can be 
partitioned into two complementary sets 
$V_\mathbf{k}^{\rm o}=X_\mathbf{k}^{\rm o}\cup Y_\mathbf{k}^{\rm o}$. Using this expression, one can 
calculate the Hamming weight of $\text{wt}(A^\prime\cdot \mathbf{k})=(l_\mathbf{k}^x+l_\mathbf{k}^y)m-2l_\mathbf{k}^xl_\mathbf{k}^y$,
where $l_\mathbf{k}^x=|X_\mathbf{k}^{\rm o}|$ and $l_\mathbf{k}^y=|Y_\mathbf{k}^{\rm o}|$. With this in hand, one
obtains $\text{wt}[(A^\prime\cdot \mathbf{k}) \vee \mathbf{k})] < m$ only if $l_\mathbf{k}^x=l_\mathbf{k}^y=0$ by a simple 
algebraic argument, i.e.\ all the vertices in subgraph $(K_{m,m})_\mathbf{k}$ has even 
degree.

According to Euler's theorem, every bitstring in $Z\left(L(K_{m,m}),n,m\right)$ 
corresponds to an Eulerian cycle or sum (edge symmetric difference) of Eulerian 
cycles. Furthermore, because of the special structure of $K_{m,m}$, every cycle 
in $K_{m,m}$ can be decomposed into the sum of 4-edge cycles so that a maximum 
independent subset of $Z(m)$ are chosen where each bitstring corresponds to 
a 4-edge cycle. Proving that $\mathbf{s}^v$ is orthogonal to every bitstring 
corresponding to $4$-edge cycles, one obtains the set $\mathbf{s}^v\in Z^\perp(m)$ for 
every $v\in V$. Finally, one can prove $\mathbf{s}^v\notin W(m)$ by some algebraic 
arguments, so that $\mathbf{s}^v\in C(m)$ for every $v\in V$.
\end{proof}

For the graph state represented by the line graph of the complete bipartite 
graph, the following result holds from Lemma~\ref{lemma:CGnd_qecc} and the fact 
that $\mathbf{s}^{v}\in C(L(K_{m,m}), m^2,m)$:
\begin{corollary}
The subspace 
$$\operatorname{span}_{\mathbb{C}} \Set{\Ket{L(K_{m,m})},
\Ket{\mathbf{s}^{v}}_{L(K_{m,m})}}$$
is a $[m^2,1,m]$ QECC.
\end{corollary}
The line graph $L(K_{m,m})$ resembles $L(K_m)$ in this respect, and the 
corresponding graph state suffers a similar expansion problem; it is not 
possible to expand this code without decreasing the code distance.

\section{3D toric graph code}
\label{sec:3DToircGraphCode}
In this section we first present the structure of a generalized toric graph. 
A family of $[[n,n^{1/3},n^{1/3}]]$ QECCs with geometrically local stabilizers 
is then obtained. This is considered as a generalization of toric code, and 
is referred as a 3D toric graph code in this work.

In the toric code, qubits are divided into two complementary subsets $X\cup Y$, 
corresponding to their placement on horizontal and vertical edges of the
two-dimensional lattice, respectively. In the associated toric graph, the 
induced subgraphs in $X$ and $Y$ separately are comprised of multiple star 
graphs, while the subgraph connecting $X$ and $Y$ are half graphs. One can 
consider the toric graph to have a two-layer structure: each layer is an 
$L^2$-vertex multi-star graph, and different layers are connected by 
half graphs. It is natural to extend this two-layer structure to $L$ layers. In
this way one obtains an $L^3$-vertex generalized toric graph $G_{\rm gtoric }$.
Vertices are denoted by $(i,j,k)\in [L]^3$, where $i,j,k\in[L]$. 

The adjacency matrix of the generalized toric graph $G_{\rm gtoric}$ is a 
straightforward generalization of Eq.~(\ref{eq:toric_adjacency_matrix}):
\begin{equation}
\label{eq:adjacency_3d_toric}
\begin{aligned}
&A_{i_1j_1k_1,i_2j_2k_2}\\ 
=&\delta_{j_1,j_2}\delta_{k_1,k_2}\left(\delta_{i_1,1}\theta_{2,i_2}+ \delta_{i_2,1}\theta_{2,i_1}   \right)  \\
+&\delta_{j_1j_2} (\delta_{k_1,k_2+1}\theta_{i_2,i_1}\theta_{2,i_2}
+\delta_{k_2,k_1+1}\theta_{i_1,i_2}\theta_{2,i_1} )\\
+&\delta_{j_1,j_2+1}\delta_{k_1,k_2+1}\theta_{i_2,i_1}\theta_{2,i_2}\\
+&\delta_{j_2,j_1+1}\delta_{k_2,k_1+1}\theta_{i_1,i_2}\theta_{2,i_1},
\end{aligned}
\end{equation}
where  $i_1,i_2,j_1,j_2,k_1,k_2\in [L]$. Similar to the $2L^2$-vertex toric 
graph, the vertices in $G_{\rm gtoric}$ can be partitioned into $L$ disjoint 
subsets $X_1\cup\cdots\cup X_L$ of the same size $L^2$, where
$X_i=[L]\times[L]\times\{i\}$. The induced subgraphs on $X_i$ are also composed 
of multiple star graphs and $X_i$ and $X_{i+1}$ are connected by half graphs, 
similar to $X$ and $Y$ in the toric graph.
\begin{theorem}
\label{theorem:3d_toric_TQO1}
The family of generalized toric graph state $\mathtt{S}=\Set{\Ket{G_{\rm gtoric}} \in\mathscr{H}_2^{\otimes L^3} }_{L\in\mathbb{N}}$ is in class TQO-1.
\end{theorem}
The graph state $\Ket{G_{\rm gtoric}}\in\mathscr{H}_2^{\otimes L^3}$ is 
topologically ordered, as stated in Theorem~\ref{theorem:3d_toric_TQO1}. The
proof that $\Set{\Ket{G_{\rm gtoric}}}$ is in TQO-1 takes a different approach 
than was pursued in previous sections (i.e.\ via showing that 
$d^{\rm max}\in\operatorname{sunlin}(n)$). Rather, $G_{\rm gtoric}$ is a 
generalization of the toric graph, and recall that the toric graph state is 
LC-equivalent to one of the ground states in 2D toric code. We show that it is
possible to construct a generalized toric code such that~$\Ket{G_{\rm gtoric}}$ 
is (or is LC-equivalent to) one of the states in this code. Next, we present the construction of  an 
$[[L^3,L,L]]$ QECC with {growing distance}, a large number of logical qubits, and 
local stabilizer generators, using the generalized toric graph state.

From the adjacency matrix Eq.~(\ref{eq:adjacency_3d_toric}), one readily 
obtains the stabilizer generators of the generalized toric graph state 
$\Ket{G_{\rm gtoric}}$:
\begin{eqnarray}
S_{1jk}&:=& X_{1jk}\prod_{l=2}^{L} Z_{ljk};\\
S_{ijk}& := &X_{ijk}Z_{1jk} \left(\prod_{l=i}^L Z_{lj,k+1} \right) 
\left(\prod_{l=2}^i Z_{lj,k-1} \right)\nonumber \\
&\times&\left(\prod_{l=i}^L Z_{l,j+1,k+1} \right) 
\left(\prod_{l=2}^i Z_{l,j-1,k-1} \right), i\ge 2,\hphantom{aa}
\end{eqnarray} 
which act on $L$ and $2L$ qubits, respectively.
First, transform the non-local stabilizer generators of~$\Ket{G_{\rm gtoric}}$ 
to a set of local but not all independent stabilizers. Let 
$B=\set{1}\times[L]\times[L]$ denote the subset of qubits labeled by $i=1$ and 
$H_B$ denote the Hadamard gate acting on the qubits in $B$:
\begin{equation}
H_B = \prod_{j,k\in [L]} H_{1jk}.    
\end{equation}

\begin{lemma}
The state $H_B\Ket{G_{\rm gtoric}}$ is stabilized by  
\begin{align}
\label{eq:S''_ijk}
S_{ijk}^{\prime\prime} = &X_{ijk}X_{i+1,j,k}Z_{i,j,k+1}Z_{i,j+1,k+1} \nonumber\\
&\times Z_{i+1,j,k-1}Z_{i+1,j-1,k-1}~{\rm mod}~L
\end{align} 
for all $i,j,k\in[L]$.
\end{lemma}
\begin{proof}

Multiplying neighboring stabilizer generators, one obtains a new set of local 
stabilizer generators:
\begin{itemize}
\item $2\le i\le L-1$:
\begin{align}
&S_{ijk}^\prime = S_{ijk}S_{i+1,jk}\\
=&X_{ijk}X_{i+1,j,k}Z_{i,j,k+1}Z_{i,j+1,k+1}Z_{i+1,j,k-1}Z_{i+1,j-1,k-1}; \nonumber
\end{align} 

\item $i=1$:
\begin{align}
S_{1jk}^\prime =& S_{2jk}S_{1,j,k+1}S_{1,j+1,k+1}\\
=&X_{2jk}Z_{1jk}Z_{2,j-1,k-1}Z_{2,j,k-1}X_{1,j,k+1}X_{1,j+1,k+1}; \nonumber
\end{align} 

\item $i=L$:
\begin{align}
S_{Ljk}^\prime =& S_{Ljk}S_{1,j,k-1}S_{1,j-1,k-1}\\
=&X_{Ljk}Z_{1jk}Z_{L,j,k+1}Z_{L,j+1,k+1} X_{1,j,k-1}X_{1,j-1,k-1}. \nonumber
\end{align}
\end{itemize}
These $S^\prime_{ijk}$ stabilize the state~$\Ket{G_{\rm gtoric}}$, but they are
not all expressed as the same combination of Pauli operators. Because $HXH=Z$ 
and $HZH=X$, applying the Hadamard conjugation 
$S^{\prime\prime}_{ijk}=H_BS^\prime_{ijk}H_B$ one obtains
\begin{align}
S_{1jk}^{\prime\prime}=& X_{1jk}X_{2jk}Z_{1,j,k+1}Z_{1,j+1,k+1}Z_{2,j-1,k-1}Z_{2,j,k-1},  \\
S_{Ljk}^{\prime\prime}=& X_{Ljk}X_{1jk}Z_{L,j,k+1}Z_{L,j+1,k+1} Z_{1,j,k-1}Z_{1,j-1,k-1},\\
S_{ijk}^{\prime\prime}=&S_{ijk}^{\prime}, 2\le i\le L-1,
\end{align}
which is {translationally  invariant},  subject to the condition that all indices of Pauli
operators are evaluated modulo $L$. 
\end{proof}

\begin{figure}[t]
     \centering
   \includegraphics[width=0.95\columnwidth]{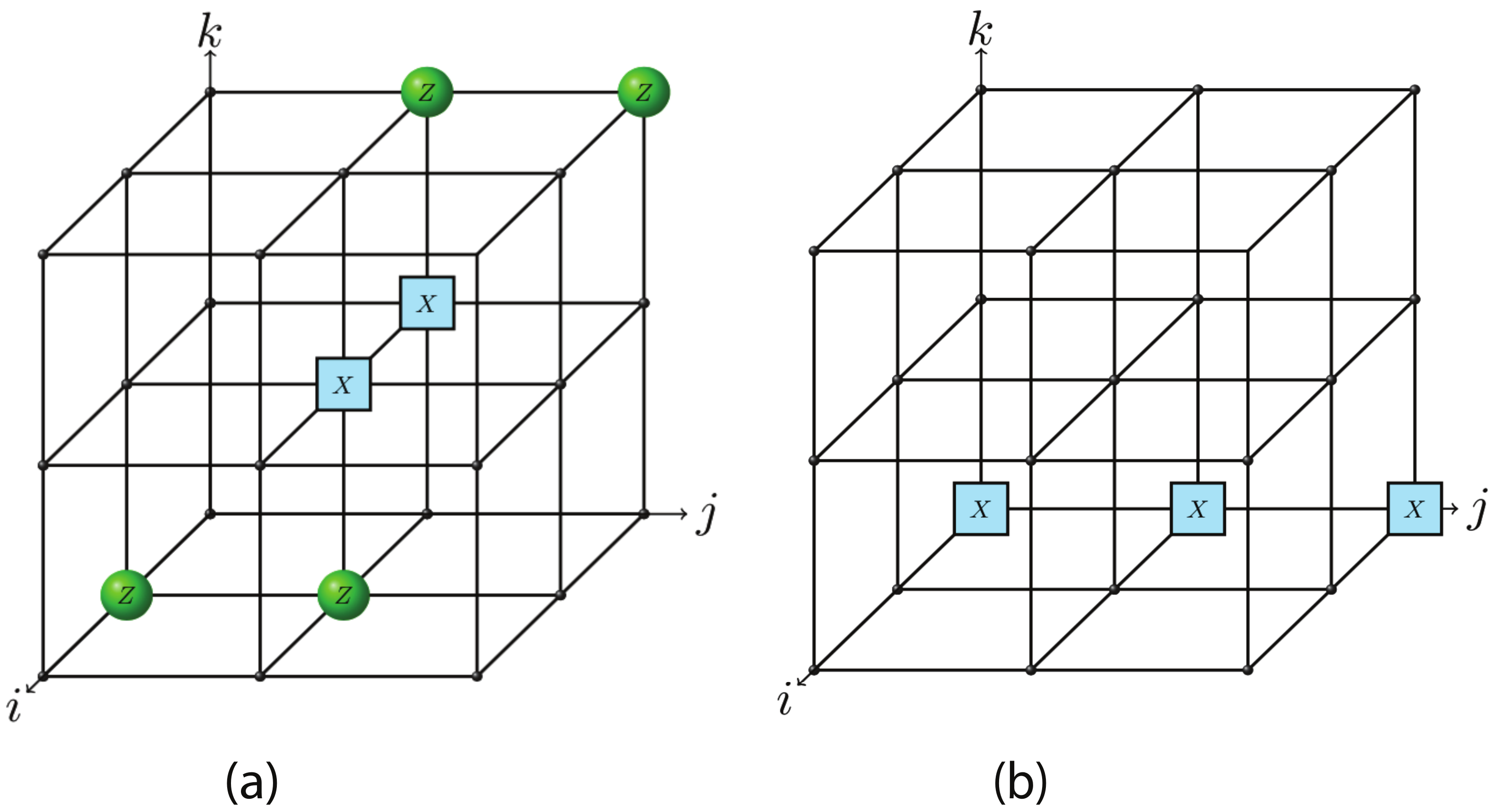}
        \caption{3D toric graph code: (a) stabilizer generator in the 3D toric 
	graph code and (b) Pauli $X$ string operator $S_1$.}
        \label{fig:3Dtoricgraphcode}
\end{figure}

Thus, the physical qubits can be arranged on the vertices of a 
three-dimensional lattice subject to periodic boundary conditions, i.e.\ a 
3-torus, and the  stabilizer generators are six-local. The geometry 
and stabilizers $S_{ijk}^{\prime\prime}$ are shown in 
Fig.~\ref{fig:3Dtoricgraphcode}(a) .
\begin{theorem}
\label{theorem:3dtoricgraphcode}
The subspace $C_{\rm T}\subseteq \mathscr{H}_2^{\otimes L^3}$ stabilized by $\{S_{ijk}^{\prime\prime}\}$ is an 
$[[L^3,L,L]]$ stabilizer QECC.
\end{theorem}
This result is rigorously proven in Appendix~\ref{appendix:3dtoricgraphcode}.
There are some known 3D codes that share a similar scaling in the distance and 
the number of logical qubits with this 3D toric graph code. For example, the 
ground-state subspace of the X-cube model~\cite{Vijay2016}, with qubits on the 
edges of an $L\times L\times L$ 3D lattice, is a $[[3L^3,6L-3,\Omega(L)]]$ 
stabilizer QECC (this scaling is in contrast with that of the 3D toric 
code~\cite{Castelnovo2008Topological} which has a similar geometry, but which is 
a $[[3L^3,3,L]]$ QECC); the Chamon model defined on a $2L\times2L\times 2L$ 
lattice is a $[[4L^3, 4L,\Omega(L)]]$ stabilizer QECC~\cite{Bravyi2011}; and the
Checkerboard model~\cite{Shirley2019} is locally equivalent to two copies of 
the X-cube model. Haah's cubic code~\cite{Haah2011local} has two qubits on each 
vertex, and has a growing number of logical qubits and macroscopic distance, 
neither of which unfortunately cannot be calculated easily.

\section{Conclusions}
\label{sec;discussion}

{In this work, we derive a set of necessary and sufficient conditions for a 
family of graph states to be topologically ordered under the definition of 
TQO-1 in Sec.~\ref{sec:topological_graph_state}. 
Using the derived criteria, we provide various pertinent examples in Sec.~\ref{sec:ExampleAnalysis};
we show that graphs with constant vertex degree and the star and complete graphs are not in 
TQO-1 in Sec.~\ref{sec:trivialgraphstate},
whereas the toric graph, multiple star graph, connected multiple star graph, the line graphs of the complete and the line graphs of complete
bipartite graphs are in TQO-1 (shown in Sec.~\ref{sec:TopoGraphCode}). 
Lastly, by generalizing the toric graph obtained from the 2D toric code, we developed 
a topological code with qubits on the vertices of a three-torus in Sec.~\ref{sec:3DToircGraphCode}, with
six-local stabilizers. The code distance and the number of 
logical qubits both scale as $n^{1/3}$, where~$n$ is the number of qubits.}

The line graphs of the complete and the line graphs of complete
bipartite graphs are chosen because they are both strongly regular graphs with vertex degree that grows with the total
number of vertices, and not because their connectivity is necessarily simple to 
generate in practice; in fact, they are unlikely to be related to stabilizer
codes with geometrically local operators. These families are relatively 
straightforward to analyze at the cost of being somewhat artificial; whether 
they are related to any known topological codes \cite{Dua2019} would is unknown.
In addition, the 3D toric graph code developed in this paper has similar behaviors to some cubic codes, such as the X-cube 
model and the Chamon model and their relation  would be an interesting avenue 
for future investigation. 

The potential of this graph-theoretic framework has not been fully exploited.
For example, is there a signature in the graph connectivity that could allow
the construction of improved LDPC codes, i.e.\ where the number of encoded 
qubits and the code distance scales better than $n^{1/3}$, where~$n$ is the 
number of physical qubits? In the current formalism there is no presumption of 
geometric locality; on the one hand this allows for the consideration of a wide 
range of graph families, but on the other could make a physical implementation 
of the code potentially daunting. Another fruitful line of inquiry is: can the 
criteria for TQO-1 developed here be used to determine if a given code is 
self-correcting? We hope to address these and related questions in future work.

\acknowledgments
This work was supported by the Natural Sciences and Engineering Research 
Council of Canada and the Major Innovation Fund of the Government of Alberta.

\appendix

\section{Stabilized subspace}
\label{appdix:stabilized_subspace}
\begin{remark*}
Given $n\in\mathbb{N}$ and two orthogonal quantum states~$\Ket{G},
\Ket{\psi}\in\mathscr{H}_2^{\otimes n}$, in which~$\Ket{G}$ is a graph state,
the subspace $\operatorname{span}_{\mathbb{C}}\{\Ket{G},\Ket{\psi}\} 
\subseteq \mathscr{H}_2^{\otimes n}$ is a stabilized subspace iff~$\Ket{\psi}$ 
is a graph basis state~$\Ket{\mathbf{h}}_G$ for some $\mathbf{h}\in\set{0,1}^n$.
\end{remark*}
\begin{proof}
If $\operatorname{span}_{\mathbb{C}}\{\Ket{G},\Ket{\psi}\} $ is a stabilized 
subspace, one can always choose $n-1$ independent and commuting stabilizers 
$\Set{S^\prime_1,\ldots,S^\prime_{n-1}}$ satisfying
\begin{align}
\label{eq:S'psi}
S^\prime_i\Ket{G}=\Ket{G}, S^\prime_i\Ket{\psi}=\Ket{\psi} \forall i\in[n-1].
\end{align}
State~$\Ket{G}$ is a stabilizer state, so there exists an element
$S_n^\prime$ independent of $\Set{S^\prime_1,\ldots,S^\prime_{n-1}}$ which
commutes with every $S_i^\prime(i\in[n-1])$, such that 
$S_n^\prime\Ket{G}=\Ket{G}$.
As commuting operators share eigenstates and the eigenvalues of $S_n^\prime$ 
are $\pm1$, either $S_n^\prime\Ket{\psi}=\Ket{\psi}$ or 
$S_n^\prime\Ket{\psi}=-\Ket{\psi}$. Because $S_n^\prime\Ket{\psi}=\Ket{\psi}$ 
leads to $\Ket{\psi}=\Ket{G}$, contradicting the fact that~$\Ket{G}$ and 
$\Ket{\psi}$ are orthogonal, one must conclude that 
$S_n^\prime\Ket{\psi}=-\Ket{\psi}$.

As the graph state~$\Ket{G}$ is also stabilized by the $\{S_i\}_{i\in[n]}$ 
from Eq.~(\ref{eq:grpah_state_S_i}), one can transform the stabilizer generator 
set $\Set{S_i^\prime}_{i\in[n]}$ to $\Set{S_i}_{i\in[n]}$ using an invertible 
matrix $R\in\mathbb{Z}_2^{n\times n}$:
\begin{equation}
S_i = \prod_{j=1}^{n}  {S_j^\prime}^{R_{ij}}, i\in[n],    
\end{equation}
where ${S_j^\prime}^0=\mathds{1}$.
With Eq.~(\ref{eq:S'psi}) and $S_n^\prime\Ket{\psi}=-\Ket{\psi}$, one obtains 
\begin{align}
S_i \Ket{\psi} 
=& \prod_{j=1}^{n} {S_j^\prime}^{R_{ij}}\Ket{\psi} =  {S_n^\prime}^{R_{in}}\Ket{\psi}\\
=& (-1)^{R_{in}}\Ket{\psi}\forall i\in[n] . \nonumber
\end{align}
The graph basis state~$\Ket{\mathbf{h}}_G$, where $\mathbf{h}\in\set{0,1}^n$ and $h_i=R_{in}$ for 
every $i\in[n]$, also satisfies
\begin{align}
\label{eq:S_i_h_G}
S_i\Ket{\mathbf{h}}_G =& X_{i}\prod_{(i,j)\in E } Z_{j}   Z_1^{\mathbf{h}_1}\otimes Z_2^{\mathbf{h}_2}\otimes\cdots \otimes Z_n^{\mathbf{h}_n}\Ket{G}  \nonumber\\
=& (-1)^{\mathbf{h}_i}Z_1^{\mathbf{h}_1}\otimes Z_2^{\mathbf{h}_2}\otimes\cdots \otimes Z_n^{\mathbf{h}_n}X_{i}\prod_{(i,j)\in E } Z_{j}   \Ket{G}  \nonumber\\
=& (-1)^{\mathbf{h}_i}\Ket{\mathbf{h}}_G =(-1)^{R_{in}}\Ket{\mathbf{h}}_G.
\end{align}
Therefore,~$\Ket{\psi}$ and~$\Ket{\mathbf{h}}_G$ are both eigenstates of $S_i$ with 
eigenvalue $(-1)^{R_{ni}}$ for every $i\in[n]$. As only one state in 
$\mathscr{H}_2^{\otimes n}$ satisfies the above condition, 
$\Ket{\mathbf{h}}_G=\Ket{\psi}$.

Furthermore, if $\Ket{\psi}=\Ket{\mathbf{h}}_G$ for some $\mathbf{h}\in\set{0,1}^n$, one can 
choose $n-1$ independent bitstrings $\{\mathbf{r}^1,\ldots,\mathbf{r}^{n-1}\}$ such that 
$\mathbf{r}^i\cdot \mathbf{h}=0$ for $i\in[n-1]$. Using $\mathbf{r}^i$, one can obtain a new set of 
stabilizers $\set{S_i^\prime}_{i\in[n-1]}$:
\begin{equation}
S_i^\prime  = \prod_{j=1}^{n} S_j^{\mathbf{r}^i_j}.    
\end{equation}
Stabilizers in $\set{S_i^\prime}_{i\in[n-1]}$ are independent because of the 
independence of the $\Set{\mathbf{r}^i}_{i\in[n-1]}$.
Then, both $S^\prime_i\Ket{G}=\Ket{G} $ and
\begin{equation}
S^\prime_i\Ket{\mathbf{h}}_G = \prod_{j=1}^{n} S_j^{\mathbf{r}^i_j} Z^\mathbf{h}\Ket{G} = (-1)^{\mathbf{r}^i\cdot\mathbf{h}} Z^\mathbf{h}S^\prime_i\Ket{G} = \Ket{\mathbf{h}}_G,   
\end{equation}
hold $\forall i\in[n-1]$ because $S_jZ_j=-Z_jS_j$ for every $j\in[n]$ and 
$S_jZ_k=Z_kS_j$ for $j\neq k$. Thus, $\operatorname{span}_{\mathbb{C}}
\Set{\Ket{G},\Ket{\mathbf{h}}_G}$ is a stabilized subspace.
\end{proof}

\section{An identity involving Pauli operators and graph basis states}
\label{appendix:graphbasisexpec}
\begin{remark*}
Given $n$-vertex graph $G$ and $\mathbf{h},\mathbf{g},\mathbf{k},\mathbf{l}\in\set{0,1}^n$, $A\in\mathbb{Z}_2^{n\times n}$  is the adjacency matrix  of $G$.
The value of ${}_G\!\braket{\mathbf{h}|X^\mathbf{k}Z^\mathbf{l}|\mathbf{g}}_G$ is
\begin{align}
{}_G\!\braket{\mathbf{h}|X^\mathbf{k}Z^\mathbf{l}|\mathbf{g}}_G =
\begin{cases}
(-1)^{\mathbf{h}\cdot \mathbf{k}+\sigma(A,\mathbf{k})},& \text{if}~A\cdot \mathbf{k}+\mathbf{l}=\mathbf{h} +\mathbf{g}, \\
0,&\text{otherwise}.
\end{cases}
\end{align}
\end{remark*}
\begin{proof}
By the definition of graph basis state in Eq.~(\ref{eq:graph_basis}) one easily obtains
\begin{align}
\label{eq:hOg}
{}_G\!\braket{\mathbf{h}|X^\mathbf{k}Z^\mathbf{l}|\mathbf{g}}_G
=&\braket{G|Z^\mathbf{h}X^\mathbf{k}Z^\mathbf{l}Z^g|G}\\
=&(-1)^{{h}\cdot{\mathbf{k}}}\braket{G|X^\mathbf{k} Z^{h+\mathbf{l}+\mathbf{g}}|G}.\nonumber
\end{align}
Because $X_i\operatorname{CZ}(i,j)= Z_j \operatorname{CZ}(i,j) X_i$, 
\begin{eqnarray}
X^\mathbf{k}\Ket{G}&=&X_1^{\mathbf{k}_1}\cdots X_n^{\mathbf{k}_n} \prod_{(i,j)\in E} 
\operatorname{CZ}(i,j) \Ket{+}^{\otimes n}\nonumber \\
&=& X_1^{\mathbf{k}_1}\cdots X_{n-1}^{\mathbf{k}_{n-1}}(Z^{A\cdot \mathbf{b}^n})^{\mathbf{k}_n}\nonumber \\
&\times&\prod_{(i,j)\in E} \operatorname{CZ}(i,j) X_n^{\mathbf{k}_n} 
\Ket{+}^{\otimes n},
\end{eqnarray}    
where $\mathbf{b}^i\in\set{0,1}^n (i\in[n])$ is the bitstring whose only non-zero
entry is in the $i$th position. Then the product $A\cdot \mathbf{b}^i\in\set{0,1}^n$ 
corresponds to the $i$th column vector of $A$. The operator 
$(Z^{A\cdot \mathbf{b}^n})^{\mathbf{k}_n}$ is $Z^{A\cdot \mathbf{b}^n}$ if $\mathbf{k}_n=1$ and is identity 
otherwise. When the $X_n$ gates are pushed through the series of 
$\operatorname{CZ}$ gates, a correction gate $Z_i$ appears on the left iff 
$(i,n)$ is a edge in graph $G$. The neighbors of $n$th vertex in graph $G$ 
correspond to the nonzero entries of the $n$th column vector of the adjacency 
matrix, so all the correction $Z$ gates can be compactly described as 
$Z^{A\cdot \mathbf{b}^n}$.

Moving $(Z^{A\cdot \mathbf{b}^n})^{\mathbf{k}_n}$ to the leftmost side, one obtains
\begin{equation}
\begin{aligned}
X^\mathbf{k}\Ket{G} =  (-1)^{\sigma_n}(Z^{A\cdot \mathbf{b}^n})^{\mathbf{k}_n} X_1^{\mathbf{k}_1}\cdots X_{n-1}^{\mathbf{k}_{n-1}}\Ket{G}
\end{aligned}
\end{equation}
where $\sigma_n =\mathbf{k}_n\cdot (\sum_{i=1}^{n-1} \mathbf{k}_i\cdot A_{in} )$.
Repeating the above procedures for different $X_i$ yields
\begin{equation}
\begin{aligned}
X^\mathbf{k}\Ket{G} 
=& (-1)^{\sum_{j=2}^n \sigma_j} (Z^{A\cdot \mathbf{b}^1})^{\mathbf{k}_1}\cdots(Z^{A\cdot \mathbf{b}^n})^{\mathbf{k}_n}\Ket{G} \\
=& (-1)^{\sum_{j=2}^n \sigma_j} Z^{A\mathbf{k}}\Ket{G},
\end{aligned}
\end{equation}
in which $\sigma_j =  \mathbf{k}_j\cdot \left(\sum_{i=1}^{j-1} \mathbf{k}_i\cdot A_{ij} \right)$ and $\sum_{j=2}^n \sigma_j=\sigma(A,\mathbf{k})$.
Therefore, the value of Eq.~(\ref{eq:hOg}) is
\begin{equation}
\begin{aligned}
{}_G\!\braket{\mathbf{h}|X^\mathbf{k}Z^\mathbf{l}|\mathbf{g}}_G
=&(-1)^{{h}\cdot{\mathbf{k}}}
\braket{G|X^\mathbf{k} Z^{\mathbf{l}+ \mathbf{h}+\mathbf{g}} |G}\\
=&(-1)^{{\mathbf{h}}\cdot{\mathbf{k}}+\sigma(A,\mathbf{k})}
\braket{G| Z^{A\mathbf{k}+\mathbf{l}+ \mathbf{h}+\mathbf{g}}|G}
\end{aligned}
\end{equation}
If $A\mathbf{k}+\mathbf{l}+ \mathbf{h} +\mathbf{g}=0^n$, then the expectation value is $(-1)^{\mathbf{h}\cdot \mathbf{k}+\sigma(A,\mathbf{k})}$;
otherwise, the result is 0.
\end{proof}

\section{Graph state in a $d$-distance QECC}
\label{appendix:QECC-C(G,n,d)}
\begin{remark*}
Given $d,n\in\mathbb{N}$ satisfying $d\le n$ and graph state $\Ket{G}\in\mathscr{H}_2^{\otimes n}$, 
$\Ket{G}$ is in an $[[n,1,d]]$ QECC iff the membership class 
$C(G,n,d)\neq\emptyset$.
\end{remark*}
\begin{proof}
Assume $\operatorname{span}\{\Ket{G},\Ket{\psi}\}$ is an $[[n,1,d]]$ QECC, 
where 
\begin{equation}
\Ket{\psi} = \sum_{\mathbf{h}\in\set{0,1}^n\backslash\{0^n\}} \alpha_{h}\Ket{\mathbf{h}}_G.     
\end{equation}
$\Ket{0^n}_G$ is excluded from the superposition because~$\Ket{G}$ and 
$\Ket{\psi}$ are orthogonal. From Corollary~\ref{coro:qcode_distance_d}, the 
following conditions are satisfied:
\begin{align}
\braket{G|X^\mathbf{k}Z^\mathbf{l}|G} =& \braket{\psi|X^\mathbf{k}Z^\mathbf{l}|\psi}, 
\forall (\mathbf{k},\mathbf{l})\in\mathcal{B}_n^{d-1}, \\
\braket{G|X^\mathbf{k}Z^\mathbf{l}|\psi} =& 0, \forall (\mathbf{k},\mathbf{l})\in\mathcal{B}_n^{d-1}.
\end{align}
Lemma~\ref{lemma:graphbasisexpec} ensures that the first condition satisfies
\begin{align}
\braket{G|X^\mathbf{k}Z^\mathbf{l}|G} =
\begin{cases}
(-1)^{\sigma(A,\mathbf{k})},& \text{if}~A\cdot \mathbf{k} =\mathbf{l}, \\
0,&\text{otherwise}.
\end{cases}
\end{align}
If $\mathbf{k}\in Z(G,n,d)$ and $l=A\cdot \mathbf{k}$, then 
$X^\mathbf{k} Z^{A\cdot \mathbf{k}} \in \mathcal{P}_n^{d-1}$ and 
$\braket{G|X^\mathbf{k}Z^{A\cdot \mathbf{k}}|G} = (-1)^{\sigma(A,\mathbf{k})}$.
Consider next the expectation value with respect to $|\psi\rangle$:
\begin{align}
&\braket{\psi|X^\mathbf{k} Z^{A\cdot \mathbf{k}}|\psi}  \\
=&\sum_{\mathbf{h},\mathbf{g}\in\set{0,1}^n\backslash\{0^n\}} \alpha_\mathbf{h}^*\alpha_g  {}_G\!\braket{\mathbf{h}|X^\mathbf{k}Z^{A\cdot \mathbf{k}}|\mathbf{g}}_G  \nonumber\\
=&\sum_{\mathbf{h},\mathbf{g}\in\set{0,1}^n\backslash\{0^n\}} \alpha_\mathbf{h}^*\alpha_g    
\begin{cases}
(-1)^{\mathbf{h}\cdot \mathbf{k}+\sigma(A,\mathbf{k})},& \text{if}~ h=g, \\
0,&\text{otherwise}.
\end{cases}\nonumber\\
=&(-1)^{\sigma(A,\mathbf{k})}\sum_{\mathbf{h}\in\set{0,1}^n\backslash\{0^n\}} |\alpha_\mathbf{h}|^2 (-1)^{\mathbf{h}\cdot \mathbf{k}}\nonumber
\end{align}
For the condition $\braket{G|X^\mathbf{k}Z^{A\cdot \mathbf{k}}|G} 
= \braket{\psi|X^\mathbf{k}Z^{A\cdot \mathbf{k}}|\psi}$ to hold when $\mathbf{k}\in Z(G,n,d)$ requires
\begin{equation}
\sum_{\mathbf{h}\in\set{0,1}^n\backslash\{0^n\}} |\alpha_\mathbf{h}|^2 (-1)^{\mathbf{h}\cdot \mathbf{k}} = 1.   
\end{equation}
But the normalization condition is
\begin{equation}
\sum_{\mathbf{h}\in\set{0,1}^n\backslash\{0^n\}} |\alpha_\mathbf{h}|^2  = 1.  
\end{equation}
Combining these two equations, one obtains
\begin{equation}
\sum_{\mathbf{h}\in\set{0,1}^n\backslash\{0^n\} \atop \mathbf{h}\cdot \mathbf{k}=1} |\alpha_\mathbf{h}|^2  = 0 
\end{equation}
and $\alpha_\mathbf{h}=0$ if there exist $\mathbf{k}\in Z(G,n,d)$ such that $\mathbf{h}\cdot \mathbf{k}=1$.
Thus, if $\alpha_\mathbf{h}\ne0$ then $\mathbf{h}\cdot \mathbf{k}=0$ for every $\mathbf{k}\in Z(G,n,d)$, 
i.e.\ $\mathbf{h}\in Z^\perp(G,n,d)$.

Finally, consider the second condition, $\braket{G|X^\mathbf{k} Z^\mathbf{l}|\psi}=0$ which 
should hold when $\text{wt}(\mathbf{k}\vee \mathbf{l})\le d-1$:
\begin{align}
&\braket{G|X^\mathbf{k} Z^\mathbf{l}|\psi}  \\
=&\sum_{\mathbf{h}\in\set{0,1}^n\backslash\{0^n\}} \alpha_\mathbf{h} \braket{G|X^\mathbf{k}Z^\mathbf{l}|\mathbf{h}}_G    \nonumber \\
=&\sum_{\mathbf{h}\in\set{0,1}^n\backslash\{0^n\}} \alpha_\mathbf{h}
\begin{cases}
(-1)^{\sigma(A,\mathbf{k})},& \text{if}~ A\cdot \mathbf{k}+\mathbf{l}=\mathbf{h}, \\
0,&\text{otherwise},
\end{cases} \nonumber \\
=&\alpha_{A\cdot \mathbf{k}+\mathbf{l}} (-1)^{\sigma(A,\mathbf{k})} = 0 \nonumber,
\end{align}
when $\text{wt}(\mathbf{k}\vee \mathbf{l})\le d-1$.
In other words, $\alpha_\mathbf{h}=0$ if there exist $\mathbf{k},\mathbf{l}\in\set{0,1}^n$ satisfying 
$\text{wt}(\mathbf{k}\vee \mathbf{l})\le d-1$ and $A\mathbf{k}+\mathbf{l}=\mathbf{h}$; more compactly, $\alpha_\mathbf{h}=0$ if 
$\mathbf{h}\in W(G,n,d)$.

In summary, if $\alpha_\mathbf{h}\ne0$ then $\mathbf{h}\in Z^{\perp}(G,n,d)$ and 
$\mathbf{h}\notin W(G,n,d)$, i.e.\ $\mathbf{h}\in C(G,n,d)$. As~$\Ket{\psi}$ is a quantum state 
orthogonal to~$\Ket{G}$, there must be at least one $\alpha_\mathbf{h}$ that is not 
zero, so the set $ C(G,n,d)$ is not empty.
\end{proof}

\section{Toric graph}
\label{appendix:ToricGraph}
Here we show that $C(G_{\rm toric}, 2L^2,L)$ is not an empty set.
Using Eq.~(\ref{eq:toric_adjacency_matrix}), one can express the action of the
adjacency matrices on the basis vectors:
\begin{align}
A\cdot \mathbf{b}^{Ljx}   =& \sum_{l=1}^{L-1} \mathbf{b}^{ljx};\quad 
A\cdot \mathbf{b}^{1jy}   = \sum_{l=2}^{L} \mathbf{b}^{ljy}; \\
A\cdot \mathbf{b}^{ijx}   =&\mathbf{b}^{Ljx}+\sum_{l=i+1}^{L} \mathbf{b}^{ljy} + \mathbf{b}^{l(j-1)y},\; i<L; \\
A\cdot \mathbf{b}^{ijy}   =&\mathbf{b}^{1jy}+\sum_{l=1}^{i-1} \mathbf{b}^{ljx} + \mathbf{b}^{l(j+1)x},\; i>1.
\end{align}
These imply the following:
\begin{align}
&A\cdot (\mathbf{b}^{ijx}+\mathbf{b}^{(i+1)jx}) =  \mathbf{b}^{(i+1)jy}+ \mathbf{b}^{(i+1)(j-1)y},\; i<L-1; 
\nonumber\\
&A\cdot \mathbf{b}^{(L-1)jx} = \mathbf{b}^{Ljx}+\mathbf{b}^{(L-1)jy} + \mathbf{b}^{(L-1)(j-1)y};\nonumber \\
&A\cdot (\mathbf{b}^{ijy}+\mathbf{b}^{(i+1)jy}) = \mathbf{b}^{ijx} +\mathbf{b}^{i(j+1)x},\; i\in[2,L-1]\nonumber \\
&A\cdot \mathbf{b}^{2jy} = \mathbf{b}^{1jy}+ \mathbf{b}^{1jx}+\mathbf{b}^{1(j+1)x};\nonumber\\
&A\cdot (\mathbf{b}^{Ljx}+\mathbf{b}^{L(j+1)x}+\mathbf{b}^{Ljy}) =\mathbf{b}^{1jy};\nonumber\\
&A\cdot (\mathbf{b}^{1jy}+\mathbf{b}^{1(j+1)y}+\mathbf{b}^{1(j+1)x}) =\mathbf{b}^{L(j+1)x}.\nonumber
\end{align}
All of the bitstrings on the left sides above are in $Z(G_{\rm toric},n,L)$,
assuming that $L$ is larger than 4. Consider the bitstrings in the first two 
rows, $\mathbf{b}^{ijx}+\mathbf{b}^{(i+1)jx}$ for $i<L-1$ and $\mathbf{b}^{(L-1)jx}$. If $\mathbf{k}\in Z^\perp(L)$ 
then $\mathbf{k}^{ijx}=0$ for every $j$ and $i<L$. Analogously, orthogonality to the 
states $\mathbf{b}^{ijy}+\mathbf{b}^{(i+1)jy},\;i\in[2,L-1]$ and $\mathbf{b}^{2jy}$ requires $\mathbf{k}^{ijy}=0$ 
for every $j$ and $i\in[2,L]$. At this point, 
$\mathbf{k}\in\operatorname{span}_{\mathbb{Z}_2}\{\mathbf{b}^{Ljx},\mathbf{b}^{1jy}\}$ for all $j$.
Moreover, because $\mathbf{b}^{Ljx}+\mathbf{b}^{L(j+1)x}+\mathbf{b}^{Ljy},\mathbf{b}^{1jy}+\mathbf{b}^{1(j+1)y}+\mathbf{b}^{1(j+1)x}
\in Z(L)$, the only possible $\mathbf{k}\in Z^\perp(L)$ are $\sum_{j=1}^{L} \mathbf{b}^{Ljx}$, 
$\sum_{i=1}^{L} \mathbf{b}^{1jy}$, and $\sum_{i=1}^{L}\left(\mathbf{b}^{Ljx}+\mathbf{b}^{1jy}\right)$.
Given that these six bitstrings comprise a maximum independent subset of 
$Z(L)$,
\begin{equation}
Z^\perp(L) = \left\{\sum_{j=1}^{L} \mathbf{b}^{Ljx},\sum_{i=1}^{L} \mathbf{b}^{1jy},\sum_{i=1}^{L} ( \mathbf{b}^{Ljx}+ \mathbf{b}^{1jy})\right\}.     
\end{equation}
The three bitstrings constituting $Z^{\perp}(L)$ are not in $W(L)$ for reasons
similar to those discussed in Sec.~(\ref{sec:multistargraph}), so they are in 
$C(L)$, and the family of toric graph states is therefore in TQO-1.

\section{Line graph of the complete graph}
\label{appendix:linegraphcompletegraph}
In this section, we present a rigorous proof of 
Theorem~\ref{theorem:line_complete_TQO1}.
\begin{remark*}
The family of line graph states $\mathtt{S}=\Set{\Ket{T_m}}_{m\ge2}$ is in 
TQO-1.
\end{remark*}
\begin{proof}
This is proven by showing that
$$C\left(T_m, \binom{m}2,\lfloor m/2\rfloor\right)\ne \emptyset.$$
As has been the case for other examples, the determination of this set requires
the analysis of which bitstrings are in the sets $Z(\lfloor m/2\rfloor)$ and 
$W(\lfloor m/2\rfloor)$, both of which involve bitstrings $A^\prime\cdot \mathbf{k}$ for 
$\mathbf{k}\in\set{0,1}^{|V^\prime|} =\set{0,1}^{|E|}$ (recall that $A'$ is the 
adjacency matrix of the line graph of $K_m$). For the complete graph 
$K_m=(V,E)$ and bitstring $\mathbf{k}\in\set{0,1}^{|E|}$, $(K_m)_\mathbf{k}=(V_\mathbf{k},E_\mathbf{k})$ is a 
subgraph of $K_m$ whose edges are labeled by the non-zero entries in $\mathbf{k}$. 
Denote the set of vertices in $V_\mathbf{k}$ with odd degree in subgraph $(K_m)_\mathbf{k}$ as 
$V_\mathbf{k}^{\rm o}$ and its size as $l_\mathbf{k}=|V_\mathbf{k}^{\rm o}|$.
\begin{lemma}
\label{lemma:weight_A'_k_com}
$\text{wt}(A^\prime\cdot \mathbf{k})=l_\mathbf{k}(m-l_\mathbf{k})$.
\end{lemma}
\begin{proof}
Let $e$ denote both an edge in $K_m$ and a vertex in $T_m$.
With $\mathbf{k}=\sum_e \mathbf{k}_e\mathbf{b}^e$, where $\mathbf{k}_e$ is the $e$th bit in $\mathbf{k}$, one obtains
$A^\prime\cdot \mathbf{k}=\sum_e \mathbf{k}_e(A^\prime \cdot \mathbf{b}^e)$, where $A^\prime \cdot \mathbf{b}^e$ is 
the $e$th column vector of $A^\prime$. 
The column vector of $A^\prime$ corresponding to the edge $e_{ij}=(v_i,v_j)$ 
only has non-zero entries at edges (not including $e$ itself) incident to 
$v_i$ or $v_j$; mathematically, 
${A^\prime} \cdot \mathbf{b}^{e_{ij}}= \mathbf{s}^{v_i} + \mathbf{s}^{v_j}$, where 
$\mathbf{s}^{v} \in\{0,1\}^{|E|}(v\in V)$ denotes the length-$n$ bitstring only having 
nonzero entries at edges incident to $v$ in $K_m$, i.e.\
\begin{equation}
\label{eq:s_v}
\mathbf{s}^{v}_{e_{ij}} = 1~\text{if}~v=v_i~\text{or}~v=v_j, 
\forall e_{ij}=(v_i,v_j)\in E.     
\end{equation}
For example, $\mathbf{s}^{v_1}$ only has non-zero entries at edges 
$\{(v_1,v_2),(v_1,v_3),\ldots,(v_1,v_m)\}$ and  $\mathbf{s}^{v_2}$ only has non-zero 
entries at edges $\{(v_2,v_1),(v_2,v_3),\ldots,(v_2,v_m)\}$. As a result, 
$\mathbf{s}^{v_1}+\mathbf{s}^{v_2}$ only has nonzero entries at 
$\{(v_1,v_3),\ldots,(v_1,v_m),(v_2,v_3),\ldots,(v_2,v_m)\}$; likewise for
$A^\prime \cdot \mathbf{b}^{e_{12}}$. Thus, $A^\prime \cdot \mathbf{b}^{e_{12}}=\mathbf{s}^{v_1}+\mathbf{s}^{v_2}$.

One therefore obtains
\begin{align}
A^\prime\cdot \mathbf{k} = \sum_{e_{ij}\in E } \mathbf{k}_{e_{ij}} (\mathbf{s}^{v_i} + \mathbf{s}^{v_j}).
\label{eq:Aprimedotk}
\end{align}
If $\mathbf{k}_{e_{ij}}=0$ then $e_{ij}$ makes no contribution to the sum, and the sum
is only over the edges where $\mathbf{k}$ has non-zero entries.
For a given ${\mathbf{k}}\in\set{0,1}^{|E|}$ and $E_\mathbf{k}$ the subset of edges 
corresponding to non-zero entries in $\mathbf{k}$, Eq.~(\ref{eq:Aprimedotk}) becomes
\begin{align}
A^\prime\cdot \mathbf{k} = \sum_{e_{ij}\in E_\mathbf{k} }  (\mathbf{s}^{v_i} + \mathbf{s}^{v_j}).
\end{align}
The sum over edges in $E_\mathbf{k}$ can also be written as a sum over all vertices in 
$V_\mathbf{k}$. The number of times $\mathbf{s}^{v_i}$ appears in the sum corresponds to the 
degree of $v_i$ in subgraph $(K_m)_\mathbf{k}$, leading to 
\begin{equation}
A^\prime\cdot \mathbf{k} = \sum_{v\in V_\mathbf{k}}\operatorname{deg}(v)_\mathbf{k} \mathbf{s}^v,
\end{equation}
where $\operatorname{deg}(v)_\mathbf{k}$ is the degree of vertex $v$ in subgraph 
$(K_m)_{\mathbf{k}}$. If $\operatorname{deg}(v)_\mathbf{k}$ is even, then 
$\operatorname{deg}(v)_\mathbf{k} \mathbf{s}^v = 0^n$; if $\operatorname{deg}(v)_\mathbf{k}$ is odd,
then  $\operatorname{deg}(v)_\mathbf{k} \mathbf{s}^v = \mathbf{s}^v$. As a result, one obtains
\begin{equation}
A^\prime\cdot \mathbf{k} = \sum_{v\in V^{\rm o}_\mathbf{k} } \mathbf{s}^v.   
\end{equation}

To calculate the Hamming weight of $A^\prime\cdot \mathbf{k}$, one must determine
whether $(A^\prime\cdot \mathbf{k})_{e_{ij}}=0$ or $(A^\prime\cdot \mathbf{k})_{e_{ij}}=1$ for 
an arbitrary edge ${e_{ij}}=(v_i,v_j)\in E$:
\begin{equation}
\label{eq:A'k_e}
(A^\prime\cdot \mathbf{k})_{e_{ij}} =   \sum_{v\in V_\mathbf{k}^{\rm o}} \mathbf{s}^v_{e_{ij}} \mod 2.   
\end{equation}
Recall that $\mathbf{s}^v_{e_{ij}}=1$ iff $v=v_i$ or $v=v_j$. There are three 
possibilities for an arbitrary edge ${e_{ij}}$:
\begin{itemize}
\item both $v_i$ and $v_j$ are in $V_\mathbf{k}^{\rm o}$:\nonumber \\
$(A^\prime\cdot \mathbf{k})_{e_{ij}} =(1+1)~\mbox{mod}~2=0$;
\item none of $v_i$ and $v_j$ are in $V_\mathbf{k}^{\rm o}$: 
$(A^\prime\cdot \mathbf{k})_{e_{ij}}=0$;
\item one of $v_i,v_j$ is in $V_\mathbf{k}^{\rm o}$ and the other is not: 
$(A^\prime\cdot \mathbf{k})_{e_{ij}}=1$. 
\end{itemize}
As there are $l_\mathbf{k}$ vertices in $V_\mathbf{k}^{\rm o}$ and $m-l_\mathbf{k}$ vertices not in 
$V_\mathbf{k}^{\rm o}$, 
$l_\mathbf{k}(m-l_\mathbf{k})$ edges satisfy the third condition and therefore
$\text{wt}(A^\prime\cdot \mathbf{k})=l_\mathbf{k}(m- l_\mathbf{k})$.
\end{proof}

\begin{lemma}
\label{lemma:even_degree_set_Z}
Defining $\operatorname{deg}(v)_\mathbf{k}$ as the degree of vertex $v$ in the subgraph
$(K_m)_\mathbf{k}$, the bitstring ${\mathbf{k}}\in\set{0,1}^{|E|}$ is in 
$Z(\lfloor m/2\rfloor)$ iff $\operatorname{deg}(v)_\mathbf{k}=even$ for all vertices in 
$(K_m)_\mathbf{k}$ and $\text{wt}(\mathbf{k})<\lfloor m/2\rfloor$. 
\end{lemma}
\begin{proof}
In every graph, the number of vertices is no greater than twice the number of 
edges, as every edge only contributes at most two distinct vertices, so that
$|V_\mathbf{k}|\le 2|E_\mathbf{k}|$ in subgraph $(K_m)_\mathbf{k}$.
When $l_\mathbf{k}=m$, $\text{wt}(A^\prime\cdot \mathbf{k})=0$ and one obtains 
\begin{align}
w\left[\left(A^\prime\cdot \mathbf{k}\right)\vee \mathbf{k}\right]&=w (\mathbf{k})=|E_\mathbf{k}|\\
&\ge|V_\mathbf{k}|/2\ge|V_\mathbf{k}^{\rm o}|/2.
\nonumber
\end{align}
As $|V_\mathbf{k}^{\rm o}|/2=\lfloor m/2\rfloor$, the bitstrings $\mathbf{k}$ 
satisfying $|V_\mathbf{k}^{\rm o}|=m$ are not in $Z(\lfloor m/2\rfloor)$. 
When $l_\mathbf{k}=0$, $\text{wt}(A^\prime\cdot \mathbf{k})=0$ and 
$\text{wt}\left[\left(A^\prime\cdot \mathbf{k}\right)\vee \mathbf{k}\right]=\text{wt}(\mathbf{k})$. For any other cases 
where $l_\mathbf{k}\ne 0$ and $l_\mathbf{k}\ne m$, $\text{wt}(A^\prime\cdot \mathbf{k})>\lfloor m/2\rfloor$ from 
Lemma~\ref{lemma:weight_A'_k_com}. Thus, ${\mathbf{k}}\in Z(\lfloor m/2\rfloor)$ iff 
$l_\mathbf{k}=0$ and $\text{wt}(\mathbf{k})<\lfloor m/2\rfloor$ so that 
$\text{wt}\left[\left(A^\prime\cdot \mathbf{k}\right)\vee \mathbf{k}\right]<\lfloor m/2\rfloor$. As $l_\mathbf{k}$ 
denotes the number vertices with odd degree in the subgraph $(K_m)_\mathbf{k}$, $l_\mathbf{k}=0$ 
means that all the vertices in $(K_m)_\mathbf{k}$ have even degree. 
\end{proof}

Euler's Theorem states that a finite connected graph has an Eulerian cycle iff 
all vertices have even degree. From Lemma~\ref{lemma:even_degree_set_Z}, if 
bitstring $\mathbf{k}\in Z(\lfloor m/2\rfloor)$, then all the vertices in subgraph 
$(K_m)_\mathbf{k}$ have even degree. As $(K_m)_\mathbf{k}$ is not necessarily connected, every 
component of $(K_m)_\mathbf{k}$ has an Eulerian cycle. Thus, every bitsring in the set 
$Z(\lfloor m/2\rfloor)$ corresponds to an Eulerian cycle or sum (symmetric 
difference) of Eulerian cycles in $K_m$. The special structure of the complete 
graph ensures that every cycle in $K_m$ can be decomposed into the sum of 
triangles.
\begin{lemma}
\label{lemma:cycletotriangle}
Given $K_m=(V,E)$ and ${\mathbf{k}}\in\set{0,1}^{|E|}$, if $(K_m)_{\mathbf{k}}$ is an Eulerian
cycle, then there exist ${\mathbf{k}}^1,\ldots, {\mathbf{k}}^i\in\set{0,1}^{|E|}$ such that 
${\mathbf{k}}^1+\cdots+{\mathbf{k}}^i ={\mathbf{k}}$ and $(K_m)_{{\mathbf{k}}^{i^\prime}}$ is a triangle for every $1\le i^\prime \le i$. 
\end{lemma}
\begin{proof}
As $(K_m)_{\mathbf{k}}$ is an Eulerian cycle, without loss of generality one can 
represent it as
\begin{equation}
\{(v_1,v_2),(v_2,v_3),\ldots,(v_{j-1},v_j),(v_j,v_1)\}.    
\end{equation}
Partition the edges in the cycle into two subsets: 
$\{(v_1,v_2),(v_j,v_1) \}$, $\{(v_2,v_3),\ldots,(v_{j-1},v_j)\}$.
Adding edge $(v_2,v_j)$ to both subsets yields bitstrings
$\mathbf{k}_1=\{(v_1,v_2),(v_2,v_j),(v_j,v_1)\}$ and 
$\mathbf{k}^{1'}=\{(v_2,v_3),\ldots,(v_{j-1},v_j),(v_j,v_2)\}$, 
$\mathbf{k}^1,\mathbf{k}^{1'}\in\set{0,1}^{|E|}$, respectively. Evidently, $(K_m)_{\mathbf{k}^1}$ is a 
triangle, and $(K_m)_{\mathbf{k}^{1'}}$ is a cycle of smaller length than $(K_m)_\mathbf{k}$
(note that $\mathbf{k}=\mathbf{k}^1+\mathbf{k}^{1'}$). Applying this procedure recursively, one 
proves the lemma.
\end{proof}

Recall that two bitstrings ${\mathbf{k}}^1,{\mathbf{k}}^2\in\set{0,1}^{|E|}$ are orthogonal to 
each other iff $(K_m)_{{\mathbf{k}}^1}$ and $(K_m)_{{\mathbf{k}}^2}$ share an even number of
common edges. To prove a bitstring $\mathbf{k}$ is in 
$Z^\perp(T_m,\binom{m}2,\lfloor m/2\rfloor)$, it is necessary to prove that
$\mathbf{k}$ is orthogonal to every bitstring in $Z(\lfloor m/2\rfloor)$.
Because of Lemma~\ref{lemma:even_degree_set_Z} and Lemma~\ref{lemma:cycletotriangle}, to prove $\mathbf{k}\in Z^\perp(\lfloor m/2\rfloor)$ it suffices to show subgraph $(K_m)_\mathbf{k}$ shares an even number of edges in common with every triangle.
It turns out that the bitstring $\mathbf{s}^v$, defined in Eq.~(\ref{eq:s_v}), satisfies
this condition.
\begin{lemma}
The bitstring ${\mathbf{s}^v}\in Z^\perp(\lfloor m/2\rfloor)$ for every $v\in V$.
\end{lemma}
\begin{proof}
The subgraph graph $(K_m)_{\mathbf{s}^v}$, composed of all edges incident to $v$, is a 
star graph on $m$ vertices where the high-degree vertex is $v$. Without loss of
generality, consider $\mathbf{s}^{v_1}$:
\begin{equation}
E_{\mathbf{s}^{v_1}}=\{(v_1,v_2),(v_1,v_3),\ldots,(v_1,v_m)\}.    
\end{equation}
All triangles in $K_m$ share an even number of edges in common with 
$(K_m)_{\mathbf{s}^{v_1}}$: if the triangle $R$ composed of edges
\begin{equation}
\{(v_a,v_b),(v_b,v_c),(v_c,v_a)\}    
\end{equation}
does not contain vertex $v_1$ then $R$ does not share any edges in common with
$(K_m)_{\mathbf{s}^{v_1}}$; whereas if $R$ does contain $v_1$, then it shares two edges 
in common with $(K_m)_{\mathbf{s}^{v_1}}$. Therefore, bitstring $\mathbf{s}^{v_1}$ (and every 
$\mathbf{s}^v$) is orthogonal to all bitstrings corresponding to triangles in $K_m$.
With Lemma~\ref{lemma:cycletotriangle}, $\mathbf{s}^v$ is also orthogonal to all 
bitstrings corresponding to cycles in $K_m$. With 
Lemma~\ref{lemma:even_degree_set_Z}, one concludes that $\mathbf{s}^v$ is orthogonal to 
every bitstring in $Z(\lfloor m/2\rfloor)$.
\end{proof}

There is one final step left to prove Theorem~\ref{theorem:line_complete_TQO1}.
\begin{lemma}
Bitstring ${\mathbf{s}^v}\notin W(T_m,\binom{m}2,\lfloor m/2\rfloor)$ for every 
$v\in V$.
\end{lemma}
\begin{proof}
Without loss of generality, assume $v=v_1$ so that 
$E_{\mathbf{s}^v} =\{(v_1,v_2),(v_1,v_3),\ldots,(v_1,v_m)\}$. It suffices to prove that
if bitstrings ${\mathbf{k}},{\mathbf{l}}\in\set{0,1}^{|E|}$ satisfy 
$A^\prime\cdot \mathbf{k}+{\mathbf{l}}={\mathbf{s}^v}$, then
\begin{equation}
\text{wt}({\mathbf{k}}\vee{\mathbf{l}}) \ge \lfloor m/2\rfloor.    
\end{equation}
If  $A^\prime\cdot \mathbf{k}$ and ${\mathbf{s}^{v}}$ share $q$ non-zero entries, i.e.\ there are 
$q$ edges $e\in E$ such that $(A^\prime\cdot \mathbf{k})_e=\mathbf{s}^v_e=1$, then 
\begin{eqnarray}
q&=&\sum_{i=2}^m (A^\prime\cdot \mathbf{k})_{(v_1,v_i)},\\
\text{wt}(\mathbf{l})&=&\text{wt}(A^\prime\cdot \mathbf{k}+{\mathbf{s}^v})= \text{wt}(A^\prime\cdot \mathbf{k})+\text{wt}(\mathbf{s}^v)-2q. \nonumber\\
&=& l_\mathbf{k}(m-l_\mathbf{k})+m-1-2q.
\end{eqnarray}
Note that $\text{wt}(\mathbf{l})=\text{wt}(-A^\prime\cdot \mathbf{k}+{\mathbf{s}^v})=\text{wt}(A^\prime\cdot \mathbf{k}+{\mathbf{s}^v})$ because
addition and subtraction are equivalent mod~2. From Eq.~(\ref{eq:A'k_e}), 
$(A^\prime\cdot \mathbf{k})_{(v_1,v_i)}=1$ for $i\in[2,m]$ iff one of $v_1$ and $v_i$ is 
in $V_\mathbf{k}^{\rm o}$ and the other is not. If $\operatorname{deg}(v_1)_\mathbf{k}$ is even, 
where $\operatorname{deg}(v)_\mathbf{k}$ as the degree of vertex $v$ in the subgraph
$(K_m)_\mathbf{k}$, then $v_1\notin V_\mathbf{k}^{\rm o}$ and $(A^\prime\cdot \mathbf{k})_{(v_1,v_i)}=1$ 
if $v_i\in V_\mathbf{k}^{\rm o}$, so  $q=l_\mathbf{k}$; on the other hand, if 
$\operatorname{deg}(v_1)_\mathbf{k}$ is odd, then $(A^\prime\cdot \mathbf{k})_{(v_1,v_i)}=1$ if 
$v_i\notin V_\mathbf{k}^{\rm o}$, so  $q=m-l_\mathbf{k}$.

Consider first the scenario where $\operatorname{deg}(v_1)_\mathbf{k}$ is even and 
$q=l_\mathbf{k}$. One then has 
\begin{equation}
\text{wt}(\mathbf{l})=l_\mathbf{k}(m-l_\mathbf{k}-2)+m-1.    
\end{equation} 
When $0\le l_\mathbf{k}\le m-2$, $\text{wt}(\mathbf{l})\ge m-1\ge \lfloor m/2\rfloor$. There is at 
least one vertex in $(K_m)_\mathbf{k}$ such that $\operatorname{deg}(v)_\mathbf{k}$ is even, 
however, in which case ${l}_k\le m-1$. When $l_\mathbf{k}=m-1$, $\text{wt}(\mathbf{l})=0$ and instead 
one needs to check the weight of $\mathbf{k}$. In this case, $l_\mathbf{k}$ is even because of 
the handshaking lemma, and therefore
\begin{equation}
\text{wt}({\mathbf{k}})=|E_\mathbf{k}|\ge |V_\mathbf{k}|/2 \ge l_\mathbf{k}/2=\frac{m-1}{2}=\lfloor m/2\rfloor.
\end{equation}
This again justifies $d=\lfloor m/2\rfloor$.
Consider the second scenario when $\operatorname{deg}(v_1)_\mathbf{k}$ is odd and 
$q=m-l_\mathbf{k}$. Then
\begin{equation}
\text{wt}(\mathbf{l})=  (l_\mathbf{k}-2)(m-l_\mathbf{k})+m-1.
\end{equation}
As there is at least one vertex such that $\operatorname{deg}(v_1)_\mathbf{k}$ is odd 
and $l_\mathbf{k}$ is even, one obtains $l_\mathbf{k}\ge 2$ and $\text{wt}(\mathbf{l})\ge m-1$, which evidently
exceeds $\lfloor m/2\rfloor$.

In summary, $\text{wt}(\mathbf{k}\vee \mathbf{l})\ge \lfloor m/2\rfloor$ if $A^\prime\cdot \mathbf{k}+\mathbf{l}=\mathbf{s}^v$,
so ${\mathbf{s}^v}\notin W(T_m,\binom{m}2,\lfloor m/2\rfloor)$.
\end{proof}
It immediately follows that $\mathbf{s}^v\in C(\lfloor m/2\rfloor)$ for every $v\in V$,
so $C(\lfloor m/2\rfloor)\ne \emptyset $ holds for every $2\le m\in \mathbb{N}$.
As the number of qubits is $n=\binom{m}2$ and the distance is 
$d_n=\lfloor m/2\rfloor=\Theta(\sqrt{n})$, together with 
Theorem~\ref{theorem:TQO1graphstate}, one obtains 
$d^{\rm max}\ge \lfloor m/2\rfloor $ for the line graph of the complete graph,
and Theorem~\ref{theorem:line_complete_TQO1} is proven.
\end{proof}

\section{Line graph of the complete bipartite graph}
\label{appendix:linecompletebigraph}
In this section we give a rigorous proof for 
Theorem~\ref{theorem:line_complete_bi_TQO1}: 
\begin{remark*}
The family of line graph states $\mathtt{S}=\Set{\Ket{L(K_{m,m})}}_{m\ge4}$ is 
in TQO-1.
\end{remark*}
The proof proceeds analogously to the proof of 
Theorem~\ref{theorem:line_complete_TQO1}, i.e.\ by showing that 
$C(L(K_{m,m}),m^2, m)\ne \emptyset$ for the line graph $L(K_{m,m})$ of the 
complete (symmetric) bipartite graph $K_{m,m}$. We analyze $A^\prime\cdot \mathbf{k}$ 
for arbitrary $\mathbf{k}\in\{0,1\}^{|E|}$; then define the sets $Z\left(m\right)$ and 
$Z^\perp\left(m\right)$; then identify specific bitstrings in 
$Z^\perp\left(m\right)$ and prove they are not in $W\left(m\right)$.

For some bitstring $\mathbf{k}\in\{0,1\}^{|E|}$, the subgraph $(K_{m,m})_\mathbf{k}=(V_\mathbf{k},E_\mathbf{k})$ 
and the degree of its vertices $\operatorname{deg}(v)_\mathbf{k}$ are defined 
analogously to the line graph of the complete graph considered previously. 
Edges in $K_{m,m}$ and vertices in $L(K_{m,m})$ are denoted by $e$.
In subgraph $(K_{m,m})_\mathbf{k}$, the vertices with odd degree are contained in the 
set $V_\mathbf{k}^{\rm o}=X_k^{\rm o}\sqcup Y_k^{\rm o}$, where $X_k^{\rm o}$ and 
$Y_k^{\rm o}$ are complementary subsets containing vertices in $X$ and $Y$, 
respectively. Their sizes are $|X_k^{\rm o}|=l_\mathbf{k}^x$ and $|Y_k^{\rm o}|=l_\mathbf{k}^y$.
\begin{lemma}
$\text{wt}(A^\prime\cdot \mathbf{k}) = m(l_\mathbf{k}^x+l_\mathbf{k}^y)-2l_\mathbf{k}^xl_\mathbf{k}^y$.
\end{lemma}
\begin{proof}
As before, $A'$ is the adjacency matrix of the line graph, and 
$A^\prime\cdot \mathbf{k} =\sum_e \mathbf{k}_e(A^\prime\cdot \mathbf{b}^e)$, where $\mathbf{k}_e$ is the 
$e$-th bit of $\mathbf{k}$ and $A^\prime\cdot \mathbf{b}^e$ is the $e$-th column vector of 
$A^\prime$. $A^\prime\cdot \mathbf{b}^e$ has non-zero entries at edges which share a 
common vertex with $e$. With $e=(v_i^x,v_j^y)$, $A^\prime\cdot \mathbf{b}^e 
= \mathbf{s}^{v_i^x} + \mathbf{s}^{v_i^y}$, where $\mathbf{s}^{v}\in\set{0,1}^{|E|}(v\in V)$ denotes the 
bitstring with non-zero entries only for edges incident to $v$ in $K_{m,m}$: 
$\mathbf{s}^v_e=1$ iff $e$ is incident to $v$ in $K_{m,m}$. For example, the bitstrings
$\mathbf{s}^{v_i^x},\mathbf{s}^{v_j^y}\in\set{0,1}^{|E|}(i,j\in[m])$ only have non-zero entries 
at $\{(v_i^x,v_1^y),\ldots,(v_i^x,v_m^y)\}$ and 
$\{(v_1^x,v_j^y),\ldots,(v_m^x,v_j^y)\}$, respectively.

As a result,
\begin{equation}
A^\prime \cdot \mathbf{k} = \sum_{e=(v_i^x,v_j^y)\in E} \mathbf{k}_e ( \mathbf{s}^{v_i^x} + \mathbf{s}^{v_j^y} ).
\end{equation}
Again, one only need sum over the edges where $\mathbf{k}_e=1$, corresponding to the 
edges in the subgraph $(K_{m,m})_\mathbf{k}$:
\begin{equation}
A^\prime \cdot \mathbf{k} = \sum_{e=(v_i^x,v_j^y)\in E_\mathbf{k}} ( \mathbf{s}^{v_i^x} + \mathbf{s}^{v_j^y} ).
\end{equation}
Rewriting the sum over edges as a sum over vertices, the number of times 
$\mathbf{s}^{v}$ appears in the sum equals the degree of $v$ in subgraph $(K_{m,m})_\mathbf{k}$:
\begin{equation}
A^\prime\cdot \mathbf{k} = \sum _{v\in E_\mathbf{k}} \operatorname{deg}(v)_\mathbf{k} \mathbf{s}^v   
\end{equation}
If $\operatorname{deg}({v})_\mathbf{k}$ is even, then 
$\operatorname{deg}(v)_\mathbf{k} \mathbf{s}^{v}=0^n$; similarly, if $\operatorname{deg}(v)_\mathbf{k}$ is 
odd, then $\operatorname{deg}(v)_\mathbf{k} \mathbf{s}^{v}=\mathbf{s}^v$. Then
\begin{equation}
A^\prime\cdot \mathbf{k} = \sum_{v\in V^{\rm o}_k } \mathbf{s}^v 
=\sum_{v_i^x\in X_k^{\rm o}}  \mathbf{s}^{v_i^x} + \sum_{v_j^y\in Y_k^{\rm o}}\mathbf{s}^{v_j^y}.
\end{equation}
For an arbitrary edge $e^\prime = (v_{i^\prime}^x, v_{j^\prime}^y)$, one has 
\begin{equation}
\label{eq:A'k_e-bi}
(A^\prime\cdot \mathbf{k})_{e^\prime} = \sum_{v_i^x\in X_k^{\rm o}}\mathbf{s}^{v_i^x}_{e^\prime}
+\sum_{v_j^y\in Y_k^{\rm o}}\mathbf{s}^{v_j^y}_{e^\prime},
\end{equation}
where $\mathbf{s}^{v_i^x}_{e^\prime} =1$ iff $i=i^\prime$ and $\mathbf{s}^{v_j^y}_{e^\prime}=1$ 
iff $j=j^\prime$. The first sum therefore equals 1 if 
$v_{i^\prime}^x \in X_k^{\rm o}$ and is 0 otherwise; the second sum equals 1 if 
$v_{i^\prime}^y \in Y_k^{\rm o}$ and is 0 otherwise. There are four possible 
cases:
\begin{itemize}
\item  $v_{i^\prime}^x\in X_k^{\rm o}$ and $v_{j^\prime}^y\in Y_k^{\rm o}$: 
$(A^\prime\cdot \mathbf{k})_{e^\prime}=1+1\mod2=0$;
\item $v_{i^\prime}^x\notin X_k^{\rm o}$ and 
$v_{j^\prime}^y\notin Y_k^{\rm o}$: $(A^\prime\cdot \mathbf{k})_{e^\prime}=0+0=0$;
\item $v_{i^\prime}^x\in X_k^{\rm o}$ and $v_{j^\prime}^y\notin Y_k^{\rm o}$: 
$(A^\prime\cdot \mathbf{k})_{e^\prime}=1$;
\item $v_{i^\prime}^x\notin X_k^{\rm o}$ and $v_{j^\prime}^y\in Y_k^{\rm o}$: 
$(A^\prime\cdot \mathbf{k})_{e^\prime}=1$. 
\end{itemize}
There are $l_\mathbf{k}^x(m-l_\mathbf{k}^y)$ edges satisfying the third condition and 
$(m-l_\mathbf{k}^x)l_\mathbf{k}^y$ edges satisfying the fourth condition, so there are
$m(l_\mathbf{k}^x+l_\mathbf{k}^y)-2l_\mathbf{k}^xl_\mathbf{k}^y$ non-zero entries in $A^\prime\cdot \mathbf{k}$.
\end{proof}

With the above result, $\text{wt}(A^\prime\cdot \mathbf{k})=0$ when $l_\mathbf{k}^x=l_\mathbf{k}^y=0$. For any 
other cases, as shown below, $\text{wt}(A^\prime{\mathbf{k}}\vee{\mathbf{k}})\ge m$ and
\begin{lemma}
\label{lemma:evendegree}
Given $\mathbf{k}\in \{0,1\}^{|E|}$, if ${\mathbf{k}}\in Z(m)$, then $\operatorname{deg}(v)_\mathbf{k}
=even$ for every vertex in subgraph $(K_{m,m})_\mathbf{k}$.
\end{lemma}
\begin{proof}
Rewrite $\text{wt}(A^\prime\cdot \mathbf{k})$ as
\begin{equation}
\text{wt}(A^\prime\cdot \mathbf{k}) = ml_\mathbf{k}^x +(m-2l_\mathbf{k}^x)l_\mathbf{k}^y.    
\end{equation}
The lower bound of $\text{wt}(A^\prime{\mathbf{k}}\vee{\mathbf{k}})$ can be determined by considering the
following different cases:
\begin{itemize}
\item when $l_\mathbf{k}^x=0$, $\text{wt}(A^\prime\cdot \mathbf{k})=ml_\mathbf{k}^y\ge m$, as long as $l_\mathbf{k}^y$ is
not also 0;
\item when $1\le l_\mathbf{k}^x\leq m/2$, $m-2l_\mathbf{k}^x\ge0$ and $\text{wt}(A^\prime\cdot \mathbf{k})
\ge ml_\mathbf{k}^x\ge m $;
\item when $m/2<l_\mathbf{k}^x\le m-1$, $m-2l_\mathbf{k}^x<0$ and $\text{wt}(A^\prime\cdot \mathbf{k}) 
\ge ml_\mathbf{k}^x +(m-2l_\mathbf{k}^x)m=(m-l_\mathbf{k}^x)m \ge m$;
\item when $l_\mathbf{k}^x=m$, then $X_k^{\rm o}=X$ and there are at least $m$ edges in 
$E_\mathbf{k}$; thus, $\text{wt}(A^\prime\cdot \mathbf{k}\vee \mathbf{k})\ge \text{wt}({\mathbf{k}})\ge m$. 
\end{itemize}
Therefore, $\text{wt}(A^\prime\cdot \mathbf{k}\vee{\mathbf{k}})\ge m$ if $l_\mathbf{k}^x+l_\mathbf{k}^y\ne 0$. 
Equivalently, if $\text{wt}(A^\prime{\mathbf{k}}\vee{\mathbf{k}})< m$ then $l^x_k+l_\mathbf{k}^y=l_\mathbf{k}=0$, i.e.\ no 
vertices in subgraph $(K_{m,m})_\mathbf{k}$ has odd degree if $\mathbf{k}\in Z(m)$.
\end{proof}
As a result, every bitstring ${\mathbf{k}}\in Z(L(K_{m,m}),m^2, m)$ corresponds to a 
subgraph $(K_{m,m})_\mathbf{k}$ in which every component is an Eulerian cycle. Because 
of the special structure of $K_{m,m}$, all the cycles in $K_{m,m}$ can be 
decomposed into the sum (symmetric difference) of 4-edge cycles.
\begin{lemma}
\label{lemma:cycleto4cycle}
If subgraph $(K_{m,m})_\mathbf{k} = (V_{{\mathbf{k}}}, E_{{\mathbf{k}}})$ in $K_{m,m}$ is a cycle and 
$|E_{{\mathbf{k}}}|>4$, then there are vectors ${\mathbf{k}}^1,\ldots,{\mathbf{k}}^i \in \set{0,1}^{|E|}$ 
such that ${\mathbf{k}}={\mathbf{k}}^1+\cdots+{\mathbf{k}}^i$ and every subgraph $G_{{\mathbf{k}}^{i^\prime}} 
(1\le i^\prime\le i)$ is a 4-edge cycle. 
\end{lemma}
\begin{proof}
If $(K_{m,m})_\mathbf{k}$ is a cycle, without loss of generality, let us denote it as 
$$\{(v_1^x,v_1^y),(v_1^y,v_2^x),\ldots,(v_j^x,v_j^y),(v_j^y,v_1^x)\}.$$
The set of edges in the cycle can be partitioned into the two subsets 
$\{(v_1^x,v_1^y),(v_1^y,v_2^x),(v_j^y,v_1^x)\}$ and 
$\{(v_2^x,v_2^y),\ldots,(v_j^x,v_j^y)\}$.
Adding the edge $(v_2^x,v_j^y)$ to both subsets, one obtains
$\{(v_1^x,v_1^y),(v_1^y,v_2^x),(v_2^x,v_j^y),(v_j^y,v_1^x)\}$ and 
$\{(v_2^x,v_2^y),\ldots,(v_j^x,v_j^y),(v_j^y,v_2^x)\}$, corresponding to 
$\mathbf{k}^1,\mathbf{k}^{1'}\in\set{0,1}^{|E|}$, respectively. Then $\mathbf{k}=\mathbf{k}^1+\mathbf{k}^{1'}$, 
$G_{\mathbf{k}^1}$ is a 4-edge cycle, and $G_{\mathbf{k}^{1'}}$ is a cycle of smaller length. 
Applying above procedure recursively, one proves the lemma.
\end{proof}

Combining Lemma~\ref{lemma:evendegree} and Lemma~\ref{lemma:cycleto4cycle}, 
then ${\mathbf{k}}\in Z^\perp(m)$ if $\mathbf{k}$ is orthogonal to every bitstring corresponding 
to 4-edge cycles in $K_{m,m}$:
\begin{lemma}
$\mathbf{s}^v\in Z^\perp(L(K_{m,m}),m^2, m)$ for every $v\in V$.
\end{lemma}
\begin{proof}
Without loss of generality, assume $v=v_1^x$ so that $(K_{m,m})_{\mathbf{s}^{v_1^x}}$ is 
an $(m+1)$-vertex star graph composed of edges
\begin{equation}
\{(v_1^x,v_1^y),(v_1^x,v_2^y),\ldots,(v_1^x,v_m^y)\}.    
\end{equation}
For an arbitrary 4-edge cycle
\begin{equation}
\{(v_a^x,v_\mathbf{b}^y),(v_\mathbf{b}^y,v_c^x),(v_c^x,v_d^y),(v_d^y,v_a^x) \}    
\end{equation}
in $K_{m,m}$, there are two vertices in $X$ and two in $Y$. If none of the two 
vertices in $X$ is $v_1^x$, then this cycle shares no edges with subgraph 
$(K_{m,m})_{\mathbf{s}^{v_1^x}}$; otherwise the cycle shares two edges. So 
$(K_{m,m})_{\mathbf{s}^{v_1^x}}$ always shares an even number of edges with every 4-edge 
cycle, and $\mathbf{s}^{v_1^x}$ is always orthogonal to bitstrings representing 4-edge 
cycles in $K_{m,m}$. Because of Lemma~\ref{lemma:cycleto4cycle}, $\mathbf{s}^{v_1^x}$ is 
also orthogonal to every bitstring in $Z(m)$, and $\mathbf{s}^{v_1^x}$ (and every $\mathbf{s}^v$) 
is in $Z^\perp(m)$.
\end{proof}

One more lemma is required to prove Theorem~\ref{theorem:line_complete_bi_TQO1}:
\begin{lemma}
$\mathbf{s}^{v}\notin W(L(L_{m,m}),m^2,m)$ for every $v\in V$.
\end{lemma}
\begin{proof}
Without loss of generality, let us assume $v=v_1^x$. It suffices to prove
that if bitstrings ${\mathbf{k}},{l}\in\set{0,1}^{|E|}$ satisfy 
$A^\prime\cdot \mathbf{k}+{l}=\mathbf{s}^{v_1^x}$, then $\text{wt}({\mathbf{k}}\vee{l}) \ge m$.

If  $A^\prime\cdot \mathbf{k}$ and $\mathbf{s}^{v_1^x}$ share $q$ non-zero entries, i.e.\ there 
are $q$ edges $e\in E$ such that $(A^\prime\cdot \mathbf{k})_e=\mathbf{s}^{v_1^x}_e=1$, then
\begin{align}
q=&\sum_{i=1}^m (A^\prime\cdot \mathbf{k})_{(v_1^x,v_i^y)},\\
\text{wt}({l})= &\text{wt}(A^\prime\cdot \mathbf{k}+\mathbf{s}^{v_1^x})= \text{wt}(A^\prime\cdot \mathbf{k})+\text{wt}(\mathbf{s}^{v_1^x})-2q
\nonumber\\
=& m(l_\mathbf{k}^x+l_\mathbf{k}^y)-2l_\mathbf{k}^xl_\mathbf{k}^y+m-2q. 
\end{align}
From Eq.~(\ref{eq:A'k_e-bi}), for $i\in[m]$, 
$(A^\prime\cdot \mathbf{k})_{(v_1^x,v_i^y)}=1$ iff one of $v_1^x$ and $v_i^y$ has odd 
degree in subgraph $(K_{m,m})_\mathbf{k}$ and the other has even degree. There are two 
scenarios: if $\operatorname{deg}(v_1^x)_\mathbf{k}$ is even, then 
$(A^\prime\cdot \mathbf{k})_{(v_1^x,v_i^y)}=1$ if $v_i^y\in Y_k^{\rm o}$, so $q=l_\mathbf{k}^y$;
on the other hand, if $\operatorname{deg}(v_1^x)_\mathbf{k}$ is odd, then 
$(A^\prime\cdot \mathbf{k})_{(v_1^x,v_i^y)}=1$ if $v_i^y\notin Y_k^{\rm o}$, so 
$q=m-l_\mathbf{k}^y$.

For both possibilities it is only necessary to analyze the case where 
$l_\mathbf{k}^x,l_\mathbf{k}^y\le m-1$; if $l_\mathbf{k}^x=m$ or $l_\mathbf{k}^y=m$, then there are at least $m$ 
edges in $E_\mathbf{k}$, in which case $\text{wt}(l\vee \mathbf{k})\ge \text{wt}(\mathbf{k})=|E_\mathbf{k}|\ge m$. First consider 
the case where $\operatorname{deg}(v_1^x)_\mathbf{k}$ is even and  
\begin{eqnarray}
\text{wt}({l})&=&m(l_\mathbf{k}^x+l_\mathbf{k}^y)-2l_\mathbf{k}^xl_\mathbf{k}^y+m-2l_\mathbf{k}^y\nonumber \\
&=& (m-2l_\mathbf{k}^y)(l_\mathbf{k}^x+1)+ml_\mathbf{k}^y.
\end{eqnarray}
Analyze $\text{wt}(\mathbf{l})$ in four different domains of $l_\mathbf{k}^y$: 
\begin{itemize}
\item when $l_\mathbf{k}^y=0$, $\text{wt}(\mathbf{l})=(l_\mathbf{k}^x+1)m\ge m$;
\item when $1\le l_\mathbf{k}^y\le m/2$, then $m-2l_\mathbf{k}^y\ge0$, so $\text{wt}({l})\ge ml_\mathbf{k}^y\ge m$;
\item when $m/2< l_\mathbf{k}^y\le m-2$, then $m-2l_\mathbf{k}^y<0$, so 
\begin{eqnarray}
\text{wt}(\mathbf{l})&\ge &(m-2l_\mathbf{k}^y)(m+1)+ml_\mathbf{k}^y\nonumber \\
&=&m^2+m-l_\mathbf{k}^y(m+2)\nonumber \\
&\ge& m^2+m-(m-2)(m+2)\nonumber \\
&=&m+4;
\end{eqnarray}
\item last, when $l_\mathbf{k}^y=m-1$, 
\begin{eqnarray}
\text{wt}(\mathbf{l})&=&(-m+2)(l_\mathbf{k}^x+1)+m(m-1)\nonumber \\
&\ge& (-m+2)m+m(m-1)=m.
\end{eqnarray}
\end{itemize}
Therefore, $\text{wt}(\mathbf{k}\vee \mathbf{l})\ge m$ when $\operatorname{deg}(v_1^x)_\mathbf{k}$ is even.
On the other hand, if $\operatorname{deg}(v_1^x)_\mathbf{k}$ is odd ($l_\mathbf{k}^x\ge 1$), one
obtains $q=m-l_\mathbf{k}^y$ and 
\begin{equation}
\text{wt}(\mathbf{l}) = (m-2l_\mathbf{k}^y)(l_\mathbf{k}^x-1)+ml_\mathbf{k}^y.    
\end{equation}
\begin{itemize}
\item When $l^y_k=0$, based on the handshaking lemma, $l_\mathbf{k}^x+l_\mathbf{k}^y=l_\mathbf{k}^x$ is 
even so $l_\mathbf{k}^x\ge2$, then $\text{wt}(\mathbf{l})=(l_\mathbf{k}^x-1)m\ge m$;
\item When $1\le l_\mathbf{k}^y\le\frac{n}{2}$, one obtains $(m-2l_\mathbf{k}^y)\ge0$, and
$\text{wt}(\mathbf{l})\ge ml_\mathbf{k}^y \ge m$;
\item when $m/2< l_\mathbf{k}^y\le m-1$, one obtains $m-2l_\mathbf{k}^y<0$, and 
\begin{eqnarray}
\text{wt}(\mathbf{l})&\ge& (m-2l_\mathbf{k}^y)(m-2)+ml_\mathbf{k}^y\nonumber \\
&=& m^2-2m-l_\mathbf{k}^y(m-4)\nonumber \\
&\ge&m^2-2m-(m-1)(m-4)\nonumber \\
&=&3m-4> m,
\end{eqnarray}
where $m\ge 4$ is assumed.
\end{itemize}
Thus, $\text{wt}(\mathbf{k}\vee \mathbf{l})\ge m$ also holds when $\operatorname{deg}(v_1^x)_\mathbf{k}$ is odd.
We have therefore proven that $A^\prime\cdot \mathbf{k}+l=\mathbf{s}^{v_1^x}$ and $\text{wt}(\mathbf{k}\vee \mathbf{l})<m$ 
cannot both be simultaneously true, i.e.\ 
$\mathbf{s}^{v_1^x}\notin W(L(L_{m,m}),m^2,m)$. One can replace $v_1^x$ with an 
arbitrary vertex in $V$ and the above argument still holds.
\end{proof}

Thus, $\mathbf{s}^{v}\in C(m)$ for every $v\in V$ when $m\ge 4$ and 
$C(L(K_{m,m}),m^2,m)\neq\emptyset$ trivially holds. In other words, for a 
rook's graph, $d^{\rm max}\ge\sqrt{n}$ and 
Theorem~\ref{theorem:line_complete_bi_TQO1} is proven.

\section{3D Toric Graph Code}
\label{appendix:3dtoricgraphcode}

\begin{remark*}
The subspace $C_{\rm T}\subseteq \mathscr{H}_2^{\otimes L^3}$ is an 
$[[L^3,L,L]]$ stabilizer QECC.
\end{remark*}
\begin{proof}
Similar to the 2D toric code, the set of stabilizers 
$\{S_{ijk}^{\prime\prime}\}$ are not entirely independent because
\begin{align}
\label{eq:prodij-Sijk=I}
T_k = \prod_{i,j=1}^L S_{ijk}^{\prime\prime}=I, k\in[L].
\end{align}
The $L$ constraints in Eq.~(\ref{eq:prodij-Sijk=I}) are independent. From the 
structure of the $S^{\prime\prime}_{ijk}$ it is straightforward to verify that 
there are no other constraints that are independent of 
Eq.~(\ref{eq:prodij-Sijk=I}).
Suppose there is a subset $D\subseteq[L]\times[L]\times[L]$ such that 
$T_D=\prod_{(i,j,k)\in D} S^{\prime\prime}_{ijk}=I$. If $(1,1,1)\in D$, then 
$X_{(1,1,1)}$ and $X_{(2,1,1)}$ are in the final product. Because there is no 
other $X$ operator in the stabilizer generator, $X_{(2,1,1)}$ must be cancelled 
by including $(2,1,1)$ in $D$. Repeating the same argument, one ends up with 
$(i,1,1)\in D$ for all $2\le i\le L$. Analogously, because there is no $Z$
operator left in the final product, $(1,i,1)\in D$ holds for all $2\le i\le L$.
Applying this argument recursively for all $(i,1,1)$ and $(1,i,1)$, where 
$2\le i\le L$, then $(i,j,1)\in D$ for all $i,j\in[L]$. Therefore, 
$T_D = T_1 T_D^\prime$ if $(1,1,1)\in D$. Repeating the same argument for other 
qubits in $D^\prime$, one obtains $T_D = \prod_{i\in \beta\subseteq[L]} T_i$.
This proves that no other constraint independent of 
Eq.~(\ref{eq:prodij-Sijk=I}) exists, so there are $L^3-L$ independent 
stabilizer generators and the dimension of the stabilized subspace is then 
$2^L$.

This implies that one can choose $L$ independent and mutually commuting 
operators, which commute with all stabilizers in Eq.~(\ref{eq:S''_ijk}). A 
simple choice for such operators are Pauli $X$ string operators along (say) the 
$j$ axis of the three-dimensional grid of qubits:
\begin{equation}
S_k = \prod_{j=1}^L X_{1jk}, 
\end{equation}
as shown in Fig.~\ref{fig:3Dtoricgraphcode}(b). All the $S_k$ commute with one
another, as they each consist of only Pauli $X$ operators, and in any case act
on different subsets of qubits. In addition, every $S_k$ also commutes with 
every $S_{ijk}^{\prime\prime}$, which can be shown as follows. Without loss of 
generality, consider $S_1$, which has support on qubits 
$Q_1:=\set{1}\times[L]\times\set{1}$. The stabilizer generators
$S_{ijk}^{\prime\prime}$ trivially commute with $S_1$ if Pauli $Z$ operators in 
$S_{ijk}^{\prime\prime}$ don't have any support on qubits $Q_1$; otherwise, one
can notice from Fig.~\ref{fig:3Dtoricgraphcode} that there are two qubits in
$Q_1$ acted on by $Z$ operators in $S_{ijk}^{\prime\prime}$. As a result, $S_1$ 
and $S_{ijk}^{\prime\prime}$ always commute. The same argument holds for other 
Pauli X string operators $S_i$ as well.

It remains to determine the distance of this error correction code, which has a 
$2^L$-dimensional encoded subspace and $L$ string $X$ operators. The code 
distance of a subspace stabilized by group $S$ equals the minimum weight of 
operators in $N(S)-S$, where $N(S)$ is the normalizer of $S$, consisting of all 
gates $U$ such that $USU^\dagger=S$. In our case $S=\{S_{ijk}^{\prime\prime}\}$ 
 and the 
$S_i$ are in $N(S)-S$ as shown above. It remains to show that the $S_i$ 
operators have the minimum weight.

All operators in $N(S)-S$ can be written as products of the $S_i$ and 
$S_{ijk}^{\prime\prime}$. The $S_{ijk}^{\prime\prime}$ each have two $X$ 
operators in the same $k$ level, expanding in the $i$ direction, so any 
$S_{ijk}^{\prime\prime}$ multiplied by $S_1$ yields either $L$ $X$ operators or 
$L+2$ $X$ operators. Continuing in the same vein, multiplying arbitrary many 
$S_{ijk}^{\prime\prime}$ by $S_1$ will not produce an operator with fewer than 
$L$ $X$ operators when $k=1$. Likewise for other string operators. Thus, the 
$S_i$ are indeed the operators with minimum weight in $N(S)-S$. The distance of 
this generalized toric code is therefore $d=L$.
\end{proof}

\bibliographystyle{apsrev.bst}
\bibliography{ref}
\end{document}